%% file: ms_vbraito_3.tex
\documentclass[useAMS,usenatbib]{mn2e}
\usepackage{float,graphics,epsfig,array,amsmath,amssymb}
\usepackage{times}
\usepackage{color}

\newcommand{\flux}{{erg~cm$^{-2}$~s$^{-1}$}}
\newcommand{\ltsima}{$\; \buildrel < \over \sim \;$}
\newcommand{\simlt}{\lower.5ex\hbox{\ltsima}} 
\newcommand{\gtsima}{$\; \buildrel > \over \sim \;$}
\newcommand{\simgt}{\lower.5ex\hbox{\gtsima}} 

\newcommand{\ion}[2]{\ensuremath{\mbox{#1~{\sc #2}}}}
\newcommand{\errUD}[2]{\ensuremath{^{+#1}_{-#2}}}

\newcommand{\chidof}{\ensuremath{\chi^2/\mbox{d.o.f.}}}

\newcommand{\fekb}{\mbox{Fe  K$\beta$}}

\newcommand{\nika}{\mbox{Ni K$\alpha$}}

\newcommand{\mgxi}{\ensuremath{\mbox{\ion{Mg}{xi}}}}
\newcommand{\feka}{\ensuremath{\mbox{Fe~K}\alpha}}
\newcommand{\neix}{\ensuremath{\mbox{\ion{Ne}{ix}}}}
\newcommand{\nex}{\ensuremath{\mbox{\ion{Ne}{x}}}}

\newcommand{\fexxv}{\ensuremath{\mbox{\ion{Fe}{xxv}}}}
\newcommand{\fexxvi}{\ensuremath{\mbox{\ion{Fe}{xxvi}}}}

\newcommand{\xmm}{{XMM-\emph{Newton}}}
\newcommand{\asca}{{\emph{ASCA} }}
\newcommand{\lum}{erg~s$^{-1}$}
\newcommand{\nh}{cm$^{-2}$}
\newcommand{\nhsym}{N_{\mbox{\scriptsize H}}}

\newcommand{\chandra}{{\emph{Chandra}}}

\newcommand{\suzaku}{{\emph{Suzaku}}}
\newcommand{\sax}{{\emph{BeppoSAX}}}

\newcommand{\swift}{{\emph{Swift}}}
\newcommand{\logxi}{erg cm s$^{-1}$}
\newcommand{\sorg}{NGC~4507}

\input{refs}

\title{Decoupling absorption and continuum variability in the Seyfert 2 NGC~4507}
\author[V. Braito, L. Ballo, J.~N. Reeves, A. Ptak, G. Risaliti, T.~J. Turner]{V.  Braito$^{1}$\thanks{E-mail:valentina.braito@brera.inaf.it},  L. Ballo$^{2} $, J.~N. Reeves$^{3,4}$, G. Risaliti$^{5,6}$, A. Ptak$^{7} $, T.~J. Turner $^{4}$\\
$^{1}$INAF - Osservatorio Astronomico di Brera, Via Bianchi 46 I-23807 Merate (LC), Italy\\
$^{2}$Instituto de Fõsica de Cantabria (CSIC-UC), Avda. Los Castros s/n (Edif. Juan Jorda), E-39005 Santander, Spain\\
$^{3}$Astrophysics Group, School of Physical and Geographical Sciences, Keele University, Keele, Staffordshire ST5 5BG, UK\\
$^{4}$Department of Physics, University of Maryland, Baltimore County, Baltimore, MD 21250, USA\\
$^{5}$Harvard - Smithsonian Center for Astrophysics, 60 Garden Street, Cambridge, MA 02138, USA\\
$^{6}$INAF - Osservatorio Astrofisico di Arcetri, Largo E. Fermi 5, 50125 Firenze, Italy \\
$^{7}$Goddard Space Flight Center, Greenbelt, MD 20771, USA\\
}

\begin{document}

\date{}

\pagerange{\pageref{firstpage}--\pageref{lastpage}} \pubyear{2002}

\maketitle

\label{firstpage}

\begin{abstract}

We present the results of the \suzaku\ observation of the Seyfert 2
galaxy \sorg.  This source is one of the X-ray brightest Compton-thin
Seyfert 2s and a candidate for a variable absorber. \suzaku\  caught
\sorg\ in a highly absorbed state characterised by a high column
density ($\nhsym\sim8\times10^{23}$ \nh), a strong reflected component
($R\sim 1.9$) and a high equivalent width \feka\ emission line
($EW\sim 500$ eV).  The \feka\ emission line is  unresolved at the
resolution of the \suzaku\ CCDs ($\sigma < 30$ eV  or $FWHM < 3000$ km
s$^{-1}$) and most likely originates in a distant absorber. The \fekb\  emission line 
is also clearly detected and its intensity is marginally higher than
the theoretical value for  low ionisation Fe. A comparison with
previous observations performed with \xmm\ and \sax\ reveals that the
X-ray spectral curvature changes on a timescale of a few months. We
analysed all these historical observations, with standard models as
well as with a most recent model for a toroidal reprocessor and found
that the main driver of the  observed  2--10 keV spectral
variability is a change of the line-of-sight obscuration, varying
from $\sim4\times10^{23}$ \nh\ to $\sim9\times10^{23}$ \nh.    The
primary continuum is also variable, although its photon index does not
appear to vary, while the \feka\ line and reflection component are
consistent with being constant across the observations. This suggests
the presence of a rather constant reprocessor and that the observed
line of sight $\nhsym$ variability is either due to a certain degree of
clumpiness of the putative torus or due to the presence of a second
clumpy absorber.  \\

\end{abstract}


\section{Introduction}
The main ingredient of the widely accepted Unified Model  \citep{Antonucci} of Active Galactic Nuclei (AGN)  is the
presence   along the line of sight towards type 2  (or obscured) AGN  of optically-thick material  covering a
wide solid angle.  This absorber was thought to  be rather  uniform and distributed  in a toroidal geometry, which is located   at a pc scale
distance from the  nuclear region \citep{Urry1995}  or  in the form of a
bi-conical outflow \citep{Elvis2000}. Although this  model  has  been to first order  confirmed, it is now
clear that this is only a simple scenario, which does not hold for all  AGN (\citealt{Bianchi2012,Turner2012,Turner2009}).
In particular  X-ray observations of nearby and bright AGN unveiled the co-existence  of  multiple
absorbers/reflecting mirrors  in the  central regions,  suggesting  that the absorbers could be  located
at different scales   (from within  few tens of gravitational radii  from the central  nucleus to outside the pc-scale torus)  and  could be in part inhomogeneous (\citealt{Risaliti2002,Elvis04}).   Our  view of the  inner structure of AGN has  been also  modified by the recent finding of hard excesses in type 1 AGN, which can be modelled  with the presence of a Compton-thick gas in the line-of-sight (PDS 456, \citealt{Reeves456}, 1H 0419-577, \citealt{Turner0419} and \citealt{Tatum}).\\

The  
variability of the column density of the X-ray absorbing  gas ($\nhsym$) observed in a large number of AGN revealed that a significant fraction of the absorbing medium must be clumpy. The time-scales   of these  $N_{\rm H}$ variations, which can be directly linked to the size and distance of the absorbing
clouds,   also provided   valuable  constraints on the size  and  location  of this  obscuring  material from the central accreting black
hole (see \citealt{Risaliti2002}).  Rapid  $N_{\rm H}$ variations  have been discovered on time-scales from a few
days down to a few hours for a limited but still increasing number of  obscured or type 2 AGN: NGC 1365 (\citealt{Risaliti05, Risaliti07, Risa2009a, Maiolino2010}), NGC 4388 (\citealt{Elvis04}), NGC 4151 (\citealt{Puccetti}), NGC 7582 (\citealt{Piconcelli2007,Bianchi09, Turner2000}) and UCC~4203 \citep{Risaliti10}. Some of these extreme variations   are effectively occultation events where the  column density of the absorber changes from Compton-thin ($\nhsym  < 10^{24}$ \nh) to Compton-thick ($\nhsym  > 10^{24}$ \nh).  These $N_{\rm H}$ variations  unveiled  that a significant fraction of such absorbing clouds must be located very close to the nuclear X-ray source and, more specifically, within the broad line region (BLR).\\

  However this picture   is proven only  for   those few objects,  which show extreme and rapid   $N_H$ variations. On longer time-scales (from months to years) $N_{\rm H}$ variability   is a common property in local bright
Seyfert 2 galaxies (\citealt{Risaliti2002}).  Due to  the complexity of the measurements, when a change in the spectral shape is found it is hard to distinguish between  $\nhsym$  and
photon index variations. The main open question  for some of
the detected variations is  whether  the   spectral changes are indeed due to a variable 
circum-nuclear absorber  or   are due to variability of the intrinsic
emission.   In order to remove this degeneracy, high sensitivity  and  wide spectral coverage (to determine the continuum
component) are needed. Variability of the X-ray absorbers is a common  property of AGN. Indeed,   in type 1 AGN   (see \citealt{Turner2009,Bianchi2012}) most of the observed spectral variability can  be described with changes in the covering factors and ionisation states of the inner absorbers  (NGC~4051, \citealt{Miller2010}; MCG-06-30-15, \citealt{Miller08}; Mrk766, \citealt{Miller766,Turner766}; PDS~456, \citealt{Behar}).  Furthermore, even  if rare, occultation events have been detected  in type 1 AGN: Mrk 766 (\citealt{Risaliti11}) and NGC~3516 \citep{Turner2008}, supporting the overall picture.    \\

 The hypothesis of a clumpy structure for the absorbing ``torus" has been  recently introduced in several theoretical models  (\citealt{Nenkova2002,Nenkova2008, Elitzur2006, Elitzur2012} and reference therein),  where the torus consists of several distinct clouds,  distributed in a soft-edge  torus.
These models were originally based  on   infrared  observations  \citep{Jaffe2004,Poncelet2006}, showing an apparent similarity between the IR emission of type 1 and type 2 AGN \citep{Lutz2004,Horst2006},  but are now strongly supported by the  short-term changes of the $N_{\rm H}$  of the X-ray absorbers.
As discussed by \citet{Elitzur2012},   for a ``soft-edged''  toroidal distribution of clouds, the classification  of type 1 and type 2  does not depend solely on the viewing angle;  although the probability of  a ``unobscured view" of  the AGN decreases when the line-of-sight is far from the axis,  it is non zero.   Furthermore, this model naturally accounts for $N_{\rm H}$ variability and in particular  for   occultations events due to the transition of a single cloud.\\

 NGC~4507 is one of the X-ray brightest ($F_{(2-10~\mathrm{keV})} \sim 0.6-1.3 \times 10^{-11}$ \flux), and
nearby  ($z=0.0118$) Seyfert 2 galaxies, with an  estimated unabsorbed luminosity of $3.7\times10^{43}$ \lum (\citealt{Comastri98}). It has been observed with all the major X-ray 
observatories;   Einstein (\citealt{Kriss}); {\em Ginga} (\citealt{Awaki1991});
{\em BeppoSAX} \citep{Risa2002,Dadina2007}; ASCA \citep{Turner1997,Comastri98}; 
\xmm;   \chandra\  \citep{Matt04} and {\it Rossi X-Ray Timing Explorer} ({\it RXTE};  \citealt{Rivers}). \sorg\   is   one of the brightest Seyfert 2s detected    above 10 keV with the both  BAT detector on board Swift  and INTEGRAL.   It is part of the
 58 months  BAT catalogue\footnote{http://heasarc.gsfc.nasa.gov/docs/swift/results/bs58mon/}  (\citealt{Tueller10}; Baumgartner et al. 2012 ApJS
 submitted)  and of the  INTEGRAL  AGN catalogue ($F_{(20-100 ~\mathrm{keV})}\sim 2\times 10^{-10}$\flux;  \citealt{Beckmann2009, Malizia2009, Bassani2006}).  \sorg\ was also detected in the soft  gamma-ray band with the OSSE experiment on board the Compton Gamma Ray Observatory  \citep{Bassani1995}.\\
  
All of these X-ray observations revealed a hard X-ray
spectrum typical of a Compton-thin Seyfert 2:  an X-ray continuum  characterised by  heavy
obscuration and a strong Fe K$\alpha$  line at 6.4 keV. 
 The average measured column density is $N_{\rm H}\sim 6 \times$10$^{23}$ \nh\ \citep{Risa2002}.  
A  reflection component  with a reflection fraction ranging from 0.7 to 2.0 \citep{Risa2002,Dadina2007} was measured with the \sax\ observations, while  the high energy cut-off could not be constrained. 
 A similar result has been obtained with the {\em INTEGRAL} IBIS/ISGRI data  ($R=0.6^{+1.5}_{-0.5}$; \citealt{Beckmann2009}).\\

The
soft X-ray spectrum is  also   typical  of a Compton-thin Seyfert 2, with
several   emission lines from 0.6--3~keV range, due to ionised elements from O to Si, which require the presence of  at least two photoionised
media or a single stratified medium (\citealt{Matt04}).  A comparison between the  observations performed with \sax\ and
\asca\  showed  long-term $N_{\rm H}$ variability, which changes by a factor of 2, and  also some possible
variability of the intrinsic continuum (\citealt{Risaliti2002,Risa2002}).
Altogether the emerging picture for \sorg\ is of a complex and highly   variable absorber, as seen in  other bright
  Compton-thin Seyfert 2s. 
At least two  absorbing systems are present:   a Compton-thick reprocessor, responsible for the Fe K$\alpha$  line at 6.4 keV plus
the strong  Compton reflected component detected with  \sax\ and 
{\it RXTE} (\citealt{Rivers}),  and a variable Compton-thin
absorber.   A \chandra\ observation also provided  a detection (at 99\% significance)   of  a    \fexxv\   
absorption line  (at $\sim 6.7$ keV), which   suggested the presence of an ionised absorber
\citep{Matt04}. \\

Here, we present the results of a  \suzaku\ observation (of  net exposure  $\sim 90$ ks) of \sorg\
and a comparison with previous X-ray observations of this AGN, which  shows that  below 10 keV NGC4507
alternates  from  being in a transmission to  a ``reflection'' dominated state. The  \suzaku\
observation  has been already presented in  the statistical analysis of 88   Seyfert 2 galaxies
observed by \suzaku\ \citep{Fukazawa}, investigating the properties of the Fe K complex versus the 
amount of absorption and luminosity of the sources.  \citealt{Fukazawa} report both the presence of a  high
column density absorber as well as a strong Fe K  line complex ($EW=600\pm30$ eV). 
Here we present a more
detailed analysis of the same observation, where we investigate the Fe K  line complex allowing   not only the \feka\  but also the \fekb\  line parameters ($E$, $\sigma$ and $I$)  to vary as well as a comparison with the  previous X-ray observations.  Finally, we  also 
tested  a new model for the     toroidal  reprocessor \footnote{http://www.mytorus.com/}  \citep{Mytor}. \\

 The paper is structured as follows. The   observation  and data reduction are summarised in \S~2.  In \S~3 we
present the  modelling of the broad-band spectrum obtained with \suzaku, aimed to assess the  nature of the X-ray
absorber,  the  amount of reflection and  the  iron K emission line. In \S~4 we then compare the \suzaku\  with   the  previous \xmm\ and \sax\ observations.  
Discussion and
conclusions follow in \S~5. Throughout this paper, a concordance cosmology with H$_0=71$
km s$^{-1}$ Mpc$^{-1}$, $\Omega_{\Lambda}$=0.73, and $\Omega_m$=0.27 \citep{Spergel2003} is adopted. \\ 

 \section{Observations and data reduction}
\subsection{\suzaku}
 A \suzaku\ \citep{Mitsuda07} observation of \sorg\ was performed on  20th December 2007     for a
total exposure time of about 103 ksec (over a total duration  of $\sim 178$ ksec); a summary of the observations
is  shown in Table\,\ref{table:log_observ}. 
   \suzaku\     carries on board  four co-aligned  telescopes  each with an
X-ray CCD camera  (X-ray Imaging Spectrometer; XIS \citealp{Koyama07}) at the focal plane,  and a non imaging
hard X-ray detector (HXD-PIN; \citealt{Takahashi}).   Three   XIS (XIS0, XIS2 and
XIS3)    are  front illuminated (FI), while  the  XIS1   is back illuminated; the latter has an enhanced response in the
soft X-ray band but lower  effective area at 6 keV than the XIS-FI.   At the time of this observation 
only two of the XIS-FI  were still operating\footnote{The XIS 2    failed  on November 2006}, namely the
XIS0 and XIS3. All together the XIS and the HXD PIN instruments  cover the 0.5--10 keV and 14--70 keV bands
respectively. The cleaned  XIS event files obtained from version 2 \suzaku\ pipeline processing   were  processed using HEASOFT (version v6.6.3)  and 
the \suzaku\ reduction and analysis packages   applying the standard screening for the passage through
the South Atlantic Anomaly (SAA),  elevation angles and cut-off rigidity\footnote{The screening filters all  events  within
the   SAA  as well as  with an Earth elevation angle (ELV) $ < 5\ensuremath {{}^{\circ }}$ and  Earth
day-time elevation angles (DYE\_ELV) less than $ 20\ensuremath {{}^{\circ }}$. Furthermore, we excluded also data
within  256s of the SAA  from the XIS and within 500s of the SAA for the HXD. Cut-off
rigidity (COR) criteria of $ > 8 \,\mathrm{GV}$ for the HXD data and $ > 6 \,\mathrm{GV}$ for the XIS
were used.}.  The XIS data were selected in $3 \times 3$ and
$5\times 5$ editmodes using only good events with grades 0, 2, 3, 4, 6 and filtering the  hot and flickering pixels
with the script \textit{sisclean}.   The  net exposure times are   $87.7$ ksec  for each of the XIS and  $92.9$
ksec for the HXD-PIN.\\
 
  The XIS  source spectra  were extracted from a circular region of 2.6$'$ radius  centered on the source,  while
background spectra  were extracted from two circular regions of 2.6$'$ radius  offset from the source and   the Fe 55
calibration sources, which are in two corners of CCDs.  The XIS response (rmfs) and ancillary response (arfs) files were
produced,   using the latest calibration files available, with the \textit{ftools} tasks \textit{xisrmfgen} and
\textit{xissimarfgen} respectively.  The spectra from the  two FI  CDDs (XIS 0 and XIS 3) were combined  in  a
single source spectrum (hereafter XIS--FI) after checking for consistency, while the  BI (the XIS1) spectrum  was kept separate and fitted
simultaneously.    The net 0.6--10 keV  count rates  are: $( 0.128\pm 0.001)$ counts s$^{-1}$, $( 0.131\pm  0.001)$ counts s$^{-1}$,
$( 0.128\pm 0.001)$ counts s$^{-1}$ for the  XIS 0, XIS3 and XIS1  respectively.    Data were included from 0.6--10 keV for the XIS--FI and  0.6--8.5 keV for the XIS1 chip; the difference on
the upper boundary for the XIS1 spectra is because this CCD is optimised for the  soft X-ray band.  We also
excluded the data in the 1.6--1.9 keV energy range due to calibration uncertainties. The   XIS FI (BI) source spectra were   binned  to 1024 channels and then   to a  minimum  of 100  (50) counts per
bin,   and   $\chi^2$ statistics have been used. \\

The HXD-PIN data  were reduced following the   latest \suzaku\ data reduction guide (the ABC guide Version
4.0\footnote{http://heasarc.gsfc.nasa.gov/docs/suzaku/analysis/abc/}),   and using the rev2 data, which
include all 4 cluster units.   The HXD-PIN instrument team provides the background event file (known as  the ``tuned''
background), which accounts for  the instrumental  Non X-ray Background (NXB;
\citealp{kokubun}).  The systematic uncertainty of   this ``tuned'' background model is believed to be 
$\pm$1.3\% (at the 1$\sigma$ level for a net 20
ks exposure).   We extracted the source and background spectra   using the same common good time interval,
and  corrected the source spectrum  for the detector dead time. The net exposure time  after   screening  was
92.9 ks.  We    simulated a spectrum for the cosmic X-ray background counts  adopting the form of 
\citet{Boldt} and \citet{Gruber} and the response matrix for the diffuse emission; the resulting spectrum was then added  
to the  instrumental one.\\ 
  
 \sorg\ is detected  up to 70 keV at a level of  23.6\% above the background 
corresponding to a  signal-to-noise ratio $S/N\sim 50 $. The net count rate in the
14--70 keV band is  $0.122\pm 0.002$ counts s$^{-1}$ ($\sim 11300 $ net counts).  For the spectral 
analysis the source  spectrum of \sorg\ was rebinned   in order to have a
signal-to-noise ratio of  10 in each energy bin. A
  first estimate of the  14--70 keV flux     was derived assuming a single  absorbed 
power-law ($\Gamma=1.8\pm0.3$) model. The 14--70 keV flux is about  $\sim 9.2\times 10^{-11}$\flux\ and the 
extrapolated flux in the \swift\  band (14--195 keV) is  $\sim 1.8 \times 10^{-10}$\flux,  comparable to   
the flux reported in the BAT 58 months catalog ($\sim 1.9 \times 10^{-10}$\flux,  Baumgartner et al. 2012 ApJS
submitted).\\

 \subsection{\xmm}
 \xmm\ observed \sorg\ on  the 4th January 2001   (see Table\,\ref{table:log_observ}) with a total exposure
time of about 46.2 ksec. This observation  was presented by \citet{Matt04}.  
Since we are mainly interested in  a comparison with the \suzaku\ observation and  since 
a detailed analysis has been  already published,  we focused only on the EPIC-pn data. 
 The EPIC-pn   camera  had     the medium filter applied and was   operating in Full Frame Window mode. The
\xmm\ data were processed and cleaned using the Science Analysis Software  (SAS version 10.0.2) and analysed
using standard software packages  and   the most  recent calibrations.  In order to define
the  threshold to filter for high-background time intervals,  we  extracted the 10--12 keV light curves and   filtered
out the data when the  light curve  is   2$\sigma$ above  its  mean.  This screening yields  net exposure time
(which also includes the dead-time correction)    of $\sim 32$ ksec.  The EPIC pn  source spectrum was extracted  using a
circular region of $35''$ and  background data were extracted using two circular regions with a   
radius of $30''$ each.   Response matrices and ancillary response files at the source position  were created using the
SAS tasks \textit{arfgen} and \textit{rmfgen}.  The source spectrum was then  binned to have at
least 50 counts  in each energy bin. \\   
 
 \subsection{The \swift-BAT} 
The \swift-BAT spectrum  and the latest calibration response  file (the diagonal
matrix: diagonal.rsp) were obtained  from the 58-month  survey archive; the data reduction procedure of the
eight-channel spectrum  is described in \citet{Tueller10} and Baumgartner et al. (2012 ApJS submitted). The
net   count rate in the 14--195 keV band   is  $(2.69 \pm 0.05)  \times 10^{-3}$ counts s$^{-1}$ (corresponding to  $F_{14-195 \,\mathrm{keV}} \sim 1.9 \times 10^{-10}$\flux).   \\

\begin{table*}
\caption{Summary of  the observations used: Observatory, Epoch, Instrument,  and Net exposure times.
 The  net exposure times    are
after the screening of the cleaned event files. 
\label{table:log_observ}
}

\begin{tabular}{cccc}
 \hline
 Mission &  DATE & Instrument  & T$_{\rm(net)}$ (ks)\\
 \hline

\suzaku\ & 2007-12-20   &XIS  &   87.7\\

 \suzaku\ &  2007-12-20 &   HXD-PIN & 92.7\\
   \xmm\ &  2001-01-04 &   EPIC-pn&   32.3\\
   \sax\ 1 &1997-12-26&MECS &49.79\\
   \sax\ 1 &1997-12-26&PDS &26.97\\
   \sax\ 2 &1998-07-02&MECS &31.41\\
   \sax\ 2 &1998-07-02&PDS &16.95\\
     \sax\ 3 &1999-01-13& MECS &41.33\\
  \sax\ 3 &1999-01-13&PDS &20.08\\

 \hline
\end{tabular}
\end{table*}

\section{Spectral analysis}
All the models were fit to the data using  standard software packages (\textsc{xspec} ver.
12.7.0 \citealt{xspecref}) and including   Galactic absorption   ($N_{\rm{H,Gal}}=7.23\times 10^{20}$\nh;
\citealt{Dickey}).    In the following, unless otherwise stated, fit parameters are quoted in the rest
frame of the source  and errors are at the 90\% confidence level   for one interesting
parameter ($\Delta\chi^2=2.71$).

\subsection{The  0.6--150 keV continuum}
For the analysis we fitted simultaneously   the \suzaku\ spectra  from  the XIS-FI (0.6--10 keV), the 
XIS1(0.6--8.5 keV) HXD-PIN (14--70 keV)  and the \swift-BAT spectrum (14--150 keV).  We set the cross-normalization factor  between the HXD and the XIS-FI spectra    to 1.16, as recommended for XIS nominal
observation processed after 2008 July\footnote{http://www.astro.isas.jaxa.jp/suzaku/doc/suzakumemo/suzakumemo-2007-11.pdf;\\
http://www.astro.isas.jaxa.jp/suzaku/doc/suzakumemo/suzakumemo-2008-06.pdf}  (Manabu et al.
2007; Maeda et al. 2008), while we left the cross-normalisation with the \swift-BAT spectrum free  to vary. \\ 

We initially tested    the best-fit  continuum model presented in \citet{Matt04} for the \xmm\  data, which
 was of the mathematical form:  $F(E) = \mathrm{wabs} \times (\rm{zwabs}\times \mathrm{pow1} +  \mathrm{ pexrav }+    \mathrm{pow2)}$,  where    {\textsc {pow1}}  is the   absorbed power-law, {\textsc{pexrav}}  
is the \textsc{xspec} model for a  Compton-reflected component (\citealt{pexrav}), {\textsc{pow2}} is  the     soft
scattered  power-law continuum,  which is absorbed only by the
local Galactic absorber (wabs) and    zwabs is the intrinsic absorber.  For the  {\textsc{pexrav}}  component  we allowed only   its normalisation   to vary, while we fixed the high energy cutoff to 200 keV, the amount of reflection $R=\Omega/2\pi=1$ and the inclination angle i to 30$^\circ$, as adopted  by the previous work on the \xmm\ data. \\

\begin{figure}
\begin{center}
 \resizebox{0.5\textwidth}{!}{
\rotatebox{-90}{
\includegraphics{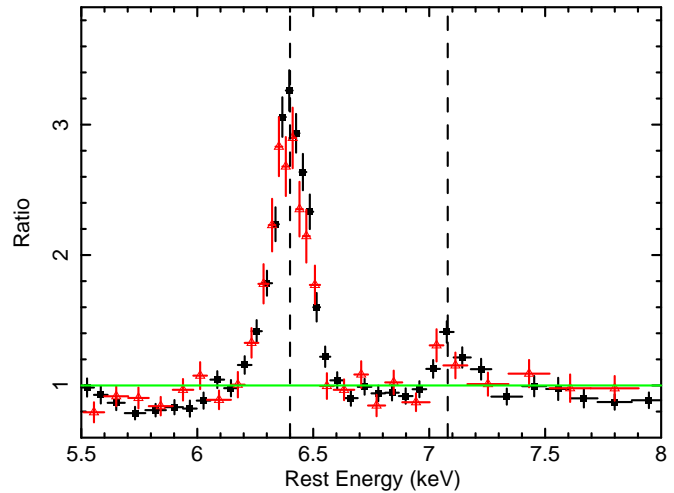}
}}
\caption{Data/model ratio between  the XIS data (XIS-FI, black  filled squares  in the electronic version; XIS-BI red  open triangles 
 in the electronic version) and the continuum model,   showing the iron line profile.  The two vertical dashed lines  correspond to the rest-frame energies of the  Fe K$\alpha$  and K$\beta$ emission lines at  6.4 keV and  7.06 keV   respectively.
\label{fig:fe_line.ps}
}
\end{center}
\end{figure}

\begin{figure}
\begin{center}
 \resizebox{0.5\textwidth}{!}{
\rotatebox{-90}{
\includegraphics{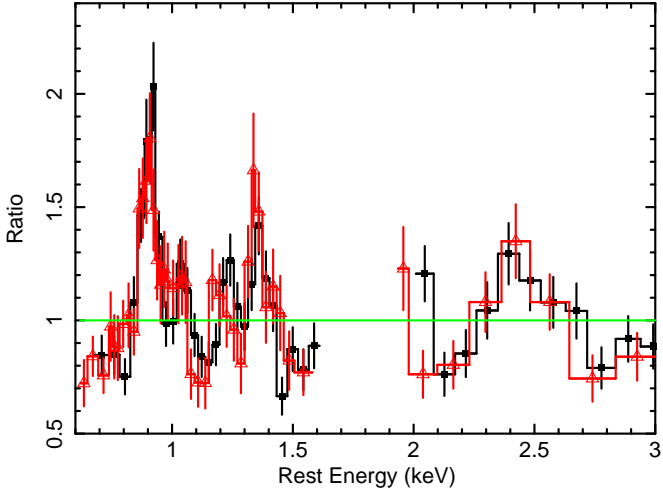}}}
\caption{Data/model ratio between  the XIS data (XIS-FI, black  filled squares  in the electronic version; XIS-BI red  open triangles 
 in the electronic version) and the continuum model (see \S~3.1),   showing the soft X-ray  lines.  
\label{fig:soft.ps}
}
\end{center}
\end{figure}

As shown in  Fig.~\ref{fig:fe_line.ps}   the residuals  to our baseline continuum
model at the  energy of the Fe K band clearly reveal the presence of a   strong
narrow core  at the expected energy of the Fe K$\alpha$ ($6.4 $ keV), as well as a  strong Fe
K$\beta$ ($\sim 7.06$ keV). We then  included  two narrow Gaussian lines  to account
for  Fe K$\alpha$ and K$\beta$ emission lines (see  \S 3.2).  Initially we fixed
the  energy   of  the  Fe K$\beta$  emission line to 7.06 keV, we tied its width   
to the width of the Fe K$\alpha$ emission line  and its normalisation to be 13.5\% 
of the Fe K$\alpha$ emission line (\citealt{Palmeri}). The inclusion of the   
lines  improves  the fit by  $\Delta\chi^2=1511$  for 3 degrees of
freedom (\chidof\  =723.7/384);   statistically  the fit is still unacceptable,  with  most of the remaining residuals being at $E< 2$ keV.   Following  the results of previous X-ray studies of \sorg, which showed a
remarkably steep X-ray emission  below 2 keV  accompanied by several soft X-ray emission
lines (\citealt{Matt04}), we then allowed the scattered component ({\sc{pow2}}) to have a different photon index
($\Gamma$; see Table \ref{tab_continuum}) with respect to the primary power law and  we added  several narrow  Gaussian
emission  lines, which  even at the \suzaku\ CCD resolution are clearly visible (see 
Fig.~\ref{fig:soft.ps}).  \\

 \begin{table*}
\caption{Summary of the  strongest soft-X-ray emission lines as detected in the \suzaku\ spectra. The energies of the lines are   quoted in the rest
frame. Fluxes and possible identifications are reported in column  2 and 3.   The observed EW are reported in
column 4 and they are  calculated against the total  observed continuum at their respective energies.
In column 5 the improvement of fit is shown with respect to the continuum model; the value for the model with no  soft
X-ray lines  is $\chidof\ =723.7/384$. Finally in column 6 we report the lab energy   for the detected  lines. \label{table:soft_lines}  } 
\begin{tabular}{cccccc}
\hline
Energy      & Flux  &ID        & EW & $\Delta \chi^2$ & E$_{\rm {Lab}}$ \\
 (keV)       & ($10^{-6}$ph cm$^{-2}$ s$^{-1}$) & &(eV) &&(keV) \\
 
    (1)   &  (2)   &  (3)    &  (4)   & (5)  & (6)  \\

\hline
     &     &      &     &    &  \\
   0.90\errUD {0.01}{0.01}& 28.6\errUD{3.3}{3.3} & \neix\ He-$\alpha$  & 107\errUD {12}{12}       & 164.2&0.905(f); 0.915(i);0.922 (r) 
   \\   
 &   &   &     & &\\

 1.03\errUD {0.01} {0.01}& 7.8\errUD {1.9}{1.9}  &  \nex\  Ly$\alpha$& 46\errUD  {11}{11}&21.5&1.022\\
  &   &   &     & &\\
 1.22 \errUD {0.01}{0.01} & 4.6\errUD {1.3}{1.3} &  \nex\ Ly$\beta$&   47\errUD {13}{13} &20.8&1.211
 (r)\\
   &   &   &     & &\\
 1.36\errUD {0.01}{0.01}&  5.5\errUD {1.1} {1.1}& \mgxi\ He$\alpha$ &76\errUD{15}{15}&69.4 & 1.331(f); 1.343(i); 1.352 (r)\\
    &   &   &     & &\\
  2.41\errUD {0.03}{0.05}& 2.0\errUD{0.9}{0.9} &  {S\,\textsc{xiv}}  K$\alpha$& 62\errUD{27}{27}&12.5& 2.411\\
     &     &      &     &    &  \\
 
 3.70\errUD {0.03}{0.03}& 2.1\errUD{0.8}{0.8} &  {Ca}   K$\alpha$& 58\errUD{20}{22}&19.3&3.69\\
     &     &      &     &    &  \\

 \hline
\end{tabular}
\end{table*}

We detected 6 strong soft X-ray emission lines   (from  Ne, Mg and S; see Table\,\ref{table:soft_lines})
and  although relatively simple,   this model  already provides  a good fit
(\chidof\ = 416.0/372).  However, we note that  the soft
power-law component is  unusually steep
($\Gamma=3.8\pm0.2$).    The steep photon index  could  be due to the presence of other weak emission
lines, which are    unresolved at the XIS-CCD  resolution and     could be related  either to the
photoionised emitters  responsible for the strongest emission lines as seen in Compton-thin Seyfert
galaxies \citep{Bianchi06,Guainazzi07}  or to the presence of  additional emission from  a
collisionally-ionised diffuse gas.   Although we investigated different  physical
scenarios (see \S \ref{soft_lines}), we note that  the different
models tested for the soft X-ray emission did not  strongly affect the results of the hard   X-ray 
energy band.

\subsection{The soft X-ray emission}
\label{soft_lines}
Although our primary aim is the analysis of the hard X-ray emission, we 
investigated both photoionised and collisionally ionised plasmas as     sources  for the soft X-ray emission lines; to this end we fitted the 0.6--150 keV spectra replacing  in turn  the    Gaussian  emission lines with either an additional thermal component ({\sc mekal} model in {\textsc{xspec}},  \citealt{Mewe85}) or  a
grid  of photoionised emission model   generated by  {\sc xstar}
\citep{xstar}, which assumes a $\Gamma\sim 2$  illuminating continuum and a
turbulence velocity of $\sigma_{\rm {v}}=100 $ km/s. \\

We found that neither a single photoionised emission model  or a  thermal component   provide an acceptable the fit  
(\chidof\ =592.3/382 and \chidof\ = 584.7/382 for    the {\sc xstar} and \textsc{mekal} component with respect to  the model with no soft X-ray lines \chidof\ =723.7/384, respectively),  and
strong residuals are present  below 2 keV;  furthermore,  for  both these models we found a steep 
$\Gamma$ for the the soft power-law component ($\Gamma=3.2\pm0.2$).   Thus we tested  for the soft X-ray
emission a composite model consisting of:  a collisionally ionised emitter, a photoionised plasma and a
soft power-law component.  We found  that this model is now a better representation of the observed  
emission (\chidof\ =510.1/380).  The  photoionised emitter has  an  ionisation parameter\footnote{The
ionisation parameter is defined  as  $\xi=$ L$_{\mathrm {ion}}$/n$_\mathrm{e}$R$^2$, where  L  is the ionising luminosity from
1--1000 Rydberg,  R is the distance of the  gas,  ne is its electron density.}  of log$\xi=1.96$\errUD {0.11}{0.07}
\logxi, while the thermal component has a temperature of $kT=0.75$ \errUD
{0.06}{0.05} keV. This model gives a total observed (i.e. corrected only for the Galactic
absorption) 0.5--2 keV flux of $\sim 4.8 \times 10^{-13}$\flux; in this energy range the relative
contribution of the collisionally  and photoionised emitters are $F_{0.5-2}^{\mathrm{COLL}}\sim  9.7 \times 10^{-14}$\flux\  and 
$F_{0.5-2}^{\mathrm{PHOT}}\sim 1.3 \times 10^{-13}$\flux\ respectively. We note that   the limited spectral resolution of the CCD spectra prevents  us from deriving definitive conclusion on the relative importance of these two
emission components, furthermore  the photon index is still relatively steep   $\Gamma=3.0\pm 0.3$,
suggesting that   there is still a possible contribution
from unresolved emission lines or a collisionally ionised emission (e. g. a  thermal component).\\ 
 Recently we obtained  a deep  ($\sim 130$  ks) \xmm-RGS observation of \sorg, which  provided  the best
soft X-ray spectrum so far for \sorg.  The properties of the soft X-ray emission are discussed in more detail in a
companion paper  describing the  \xmm-EPIC and \xmm-RGS data (\citealt{Marinucci4507}, Wang et al.  private communication). Briefly the RGS data    unveiled that the soft X-ray emission is indeed
dominated by emission lines, and that  cannot be explained with a single photoionised or thermal  component. We
stress again  that  the different models tested for the soft X-ray emission did not  strongly affect the results
of the hard   X-ray  emission, which is our primary focus in this paper.\\ 

\subsection{The Fe K emission line complex and the  high energy spectrum}
\label{suzaku_hard}

We then considered the hard X-ray emission of \sorg\,  using for the soft X-ray
emission    the simple phenomenological model  of  a scattered power-law
component and 6 Gaussian emission lines.  The spectrum  was then parametrized  with a model of the form:  $F(E) = \mathrm{wabs} \times
(\rm{zwabs}\times \mathrm{pow1} + \mathrm{pexrav}+ \mathrm{Fe \;K}\alpha+\mathrm{Fe\; }K\beta +
\mathrm{pow2 + 6\; \mathrm{GA}_\mathrm{em}})$, where the ratio of \feka\  and \fekb\  intensities was 
 initially  fixed at 13.5\%   and   GA$_\mathrm{em}$ are the soft X-ray emission lines.
    As previously described, the photon index of the scattered power-law component ({\sc{pow2}}) was left free to vary
    independently from the 
 primary power-law component ({\sc{pow1}}). This model provides a good description of the continuum  (\chidof\ =
416.0/372);  an intrinsic column density of    $N_\mathrm{H}=(8.4\pm 0.5) \times 10^{23}$
\nh\  is required,  the photon index of the primary absorbed power law   is $\Gamma=1.83\pm 0.04$ and the   intensity of 
neutral reflection component is   $N_{\mathrm{PEXRAV}}=  (1.2\pm0.1)\times 10^{-2}$ photons cm$^{-2}$s$^{-1}$ (corresponding to $R\sim 1.6$). \\

  Taking into account the high statistics of the present data and the strength of both the  Fe K  lines, we then left free to vary  the  centroid energy  and intensity of the Fe K$\beta$  line  and we found a statistically similar best-fit
(\chidof\ =409.5/370).  For the \feka\  line core we obtained
$E=6.408$\errUD{0.005}{0.004} keV, $\sigma=35\pm10$ eV (corresponding to a $ FWHM=3860\pm1100 $ km s$^{-1}$) and  $EW=490\pm 30$ eV  (with respect
to the observed continuum). For the corresponding \fekb\ we  obtained a centroid
energy of  $E=7.07\pm 0.02 $ keV and 
$I_\mathrm{FeK\beta}=9.50 \errUD{1.79}{1.77} \times 10^{-6}$photons cm$^{-2}$s$^{-1}$, which corresponds to  a $I_\mathrm{Fe\;K\alpha}/I_\mathrm{Fe\;K\beta}$   ratio  to about $\sim 17$\%.   We  also  checked the accuracy of the energy centroids and line widths using the $^{55}$Fe calibration sources located on two corners of each of the XIS chips, which 
  produce lines from Mn K$\alpha$  (K$\alpha1$   and K$\alpha 2$  at 5.899 keV
and 5.888 keV respectively). From measuring the lines in the calibration source, we find no major energy shift or residual broadening
($E=5.903\pm 0.003$ keV, $\sigma<15$  eV).    Thus  the apparent broadening of the \feka\    emission line is not due to calibration uncertainties, however,   upon the inclusion of a  possible Compton shoulder the \feka\  is unresolved (see below and  \S 4.1). \\

\begin{figure}
\begin{center}
 \resizebox{0.5\textwidth}{!}{
\rotatebox{-90}{
\includegraphics{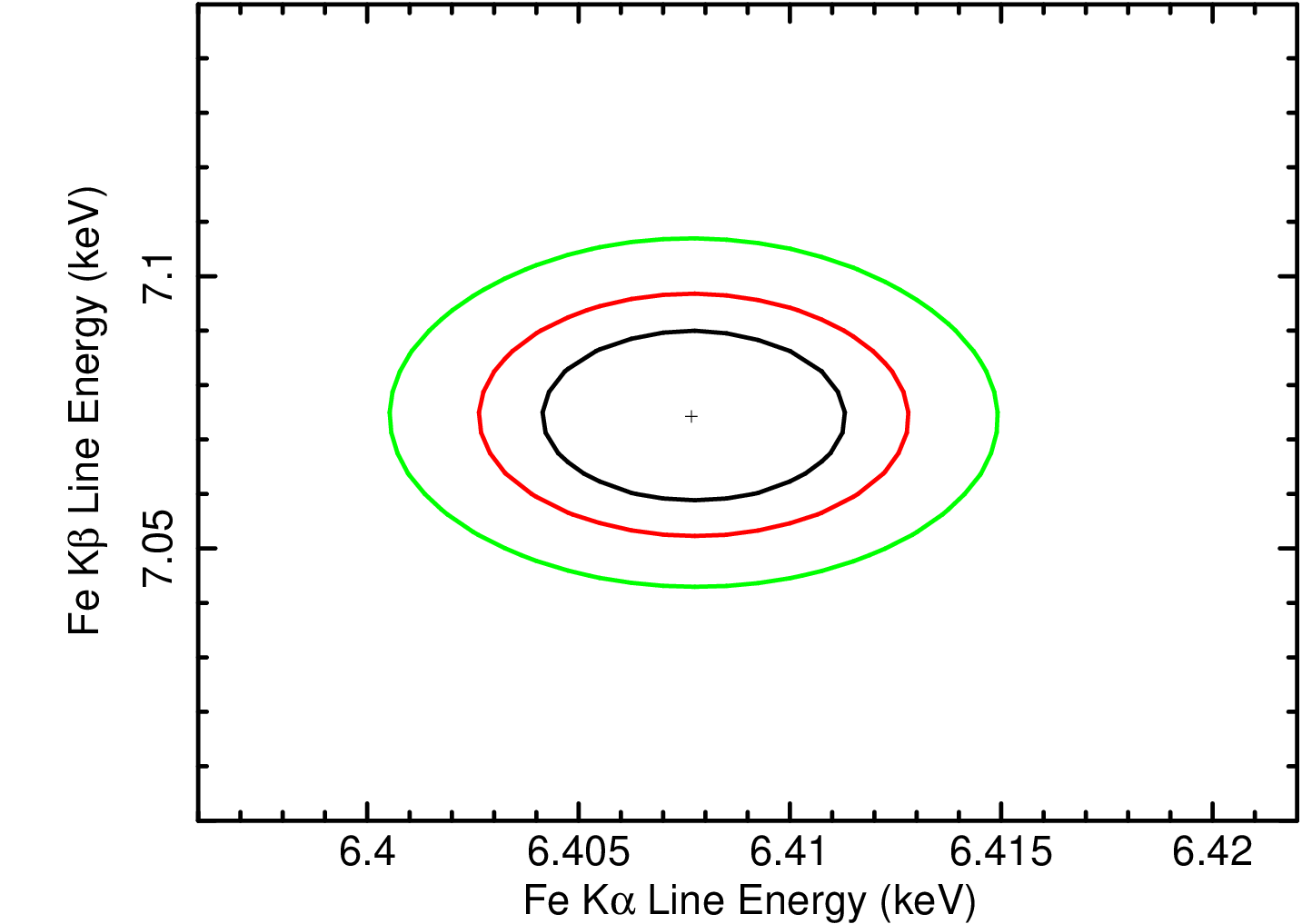}
\includegraphics{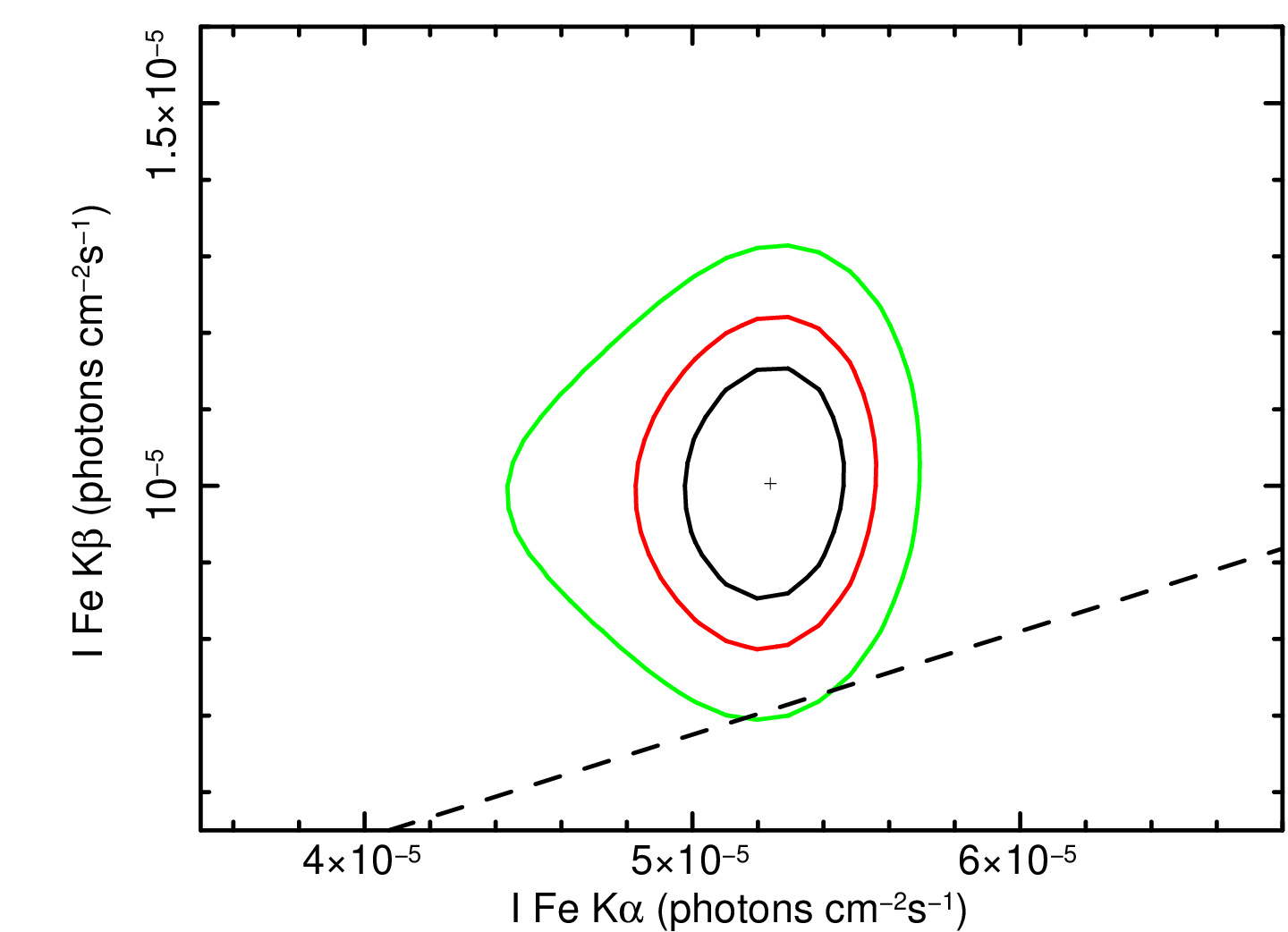}}}
\caption{Upper panel:  68\%, 90\%, and 99\% confidence contours of energy centroid of the  \feka\ versus the
\fekb\  emission lines (rest frame). Lower panel:   68\%, 90\%, and 99\% confidence contours of the intensity of  the  \feka\  versus the \fekb\ emission line; the 
 dashed line  corresponds to the \fekb/\feka\  line ratio of    0.135.
\label{fig:fe_line2.ps}
}
\end{center}
\end{figure}

The parameters derived for
the Fe  K line complex are similar to the values reported in \citet{Fukazawa}; we note
however that in our modelling the energy centroids and normalisations of all the  lines are left free to vary. In  the
upper panel of Fig.~\ref{fig:fe_line2.ps}  we show the 68\%, 90\%, and 99\% confidence contours of the narrow \fekb\ 
centroid energy versus  the centroid energy of the \feka, while   in the lower panel we show the corresponding contours
for the line intensities.    We  note that contamination to  the  \fekb\ from a possible \fexxvi\ ($E=6.97$ keV) is negligible, indeed as  can be seen in  Fig.~\ref{fig:fe_line2.ps} (upper panel) the contours of the energy centroid of the \fekb\ are fairly symmetric  and not elongated toward  lower  energies.   
Although the ratio of   \fekb\ and \feka\   intensities   is   consistent within the errors (at 99\%, see  Fig.~\ref{fig:fe_line2.ps}, lower panel) with  the 13.5\% value, as     expected for low ionisation Fe \citep{Palmeri}, it is marginally higher than the  theoretical value  for low ionisation Fe;    such a high value of the \fekb/\feka\  could be indicative that the Fe ionisation state could be as high as {\ensuremath{\mbox{\ion{Fe}{ix}}}}. In particular \citet{Palmeri} showed that while for  {\ensuremath{\mbox{\ion{Fe}{i}}}} both theoretical and experimental values of this  ratio lie in the  12--13.5\% range,   for higher ionisation states this ratio can be higher and for {\ensuremath{\mbox{\ion{Fe}{ix}}}} it can be as high as 17\%.  \\

\begin{figure}
\begin{center}
 \resizebox{0.5\textwidth}{!}{
\rotatebox{-90}{
\includegraphics{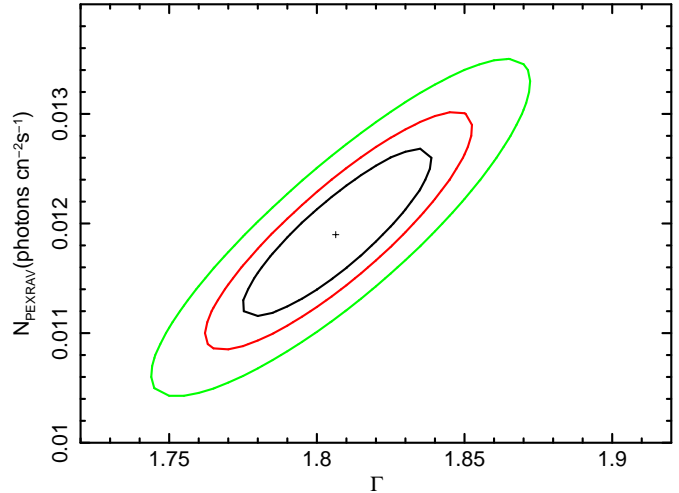}
}}
\caption{Contour plot showing the 68\%, 90\%, and 99\% confidence levels of the normalisation of the reflected
component versus $\Gamma$. Note that the contours have been produced allowing all the other parameters of the
best-fit model to vary, including the cross-normalisation  between the \suzaku\ XIS and HXD. 
\label{fig:ref.ps}
}
\end{center}
\end{figure}

\begin{figure}
\begin{center}
\resizebox{0.5\textwidth}{!}{
\rotatebox{-90}{
\includegraphics{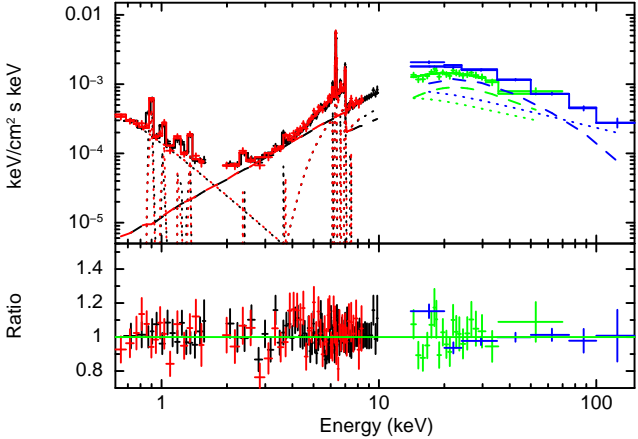}
}}
\caption{\suzaku\  and  \swift\ 0.6--150 keV   data   and best-fit model (in the electronic version: XIS-FI, black; XIS1, red; HXD-PIN  
green; \swift-BAT blue) of \sorg; data have been rebinned for plotting purposes. 
The upper panel shows the data and   model ($\Gamma\sim 1.8$; N$_\mathrm{H} \sim
8\times 10^{23}$ \nh see Table\,\ref{tab_continuum} and \S 3.3), fitted over the 0.6--150 keV  
band. The lower panel shows the data/model ratio  to this model.
 \label{fig:model1_data.ps}
}
\end{center}
\end{figure}

We note that 
the parameters  of the continuum  (see Table~\ref{tab_continuum}) and  in particular of the
reflection component are all  well constrained. In Fig.~\ref{fig:ref.ps}, we
show the confidence contours  between the normalisation of the reflection
component and the intrinsic photon index ($\Gamma$), obtained allowing   the 
 cross normalisation factor of the HXD over the XIS-FI spectrum  to 
vary.   The best fit value of the intensity of the   neutral reflection
component is   $N_{\mathrm{PEXRAV}}= 1.2\pm0.1\times 10^{-2}$ photons cm$^{-2}$
s$^{-1}$, which corresponds to $R= 1.9\pm0.2$ ;  we note that this component dominates the spectrum below $\sim 7$ keV (see Fig.~\ref{fig:model1_data.ps}).\\
 
We found also evidence for the \feka\   Compton shoulder ($E=6.23\pm 0.08$ keV), which is significant   at
the 99.8\% confidence level, according to the F-test ($\Delta\chi^{2}=11.5$ for 2 dof, \chidof\ =398.0/368). The ratio between the  Compton shoulder
intensity  and that of  the \feka\   is $\sim 9$\%, which   together with the strong
Compton reflected  component confirms the presence  of a   Compton-thick reprocessor
 \citep{Matt02,CSYaqoob}. \\

We also tested for the presence of emission lines from
\fexxv\ ($E=6.7$ keV) and \fexxvi\ ($E=6.97$ keV) adding to the model two narrow  Gaussian emission lines; 
the former  is detected at $E\sim 6.73$ keV ($\Delta\chi^{2}=8.1$, \chidof\ =389.0/366; see Table\,\ref{table:Fe_lines}) with an EW$\sim
26$ eV, while  for the \fexxvi\  emission line we can place an upper limit on its  flux to  $I_{\ion{Fe}{xxvi}} < 0.3 \times 10^{-5}
$photons cm$^{-2}$s$^{-1}$ (or $EW<26$ eV). Finally, the inspection of the residuals left by this model unveils
the presence of an emission  line like feature at  $\sim 7.5$ keV, suggesting the presence of emission from \nika. Thus we
included an additional narrow Gaussian line and we found that the parameters of this additional line are
consistent with \nika\   ($E=7.50\pm0.08$ keV, $\Delta\chi^{2}=6.7$ for 2 d.o.f.). 
 This model  now provides  a good description of the broadband X-ray spectrum of
\sorg\ (\chidof\  =383.2/364) and no strong residuals are present (see Fig.~\ref{fig:model1_data.ps}).
After the inclusion of these  additional narrow Gaussian lines the  \feka\ line is now unresolved with $\sigma<30$ eV corresponding to $FWHM < 3300 $ km s$^{-1}$.   The  \feka\   width measured with \suzaku\  is in agreement with the  \chandra\ HETG  measurement  of  $\sigma< 26 $ eV \citep{Matt04},  and suggests an origin in the torus.  We note that there  are no residuals left in the Fe K region for a possible strong  underlying broad component ($EW<120$ eV); furthermore,   leaving the width of the other emission lines free to vary does not improve the fit  ($\Delta \chi^2=1$). \\

In order to understand  if the  high value of the reflection component   could be due to the adopted model for the reflection component, we tested     the \textsc{compPS} model developed by 
\citet{Poutanen}. This model includes the processes of thermal Comptonisation of the reflected component, which is not included in the \textsc{pexrav} model. We tested both a  slab and a spherical geometry  for the reflector. We found that both these models provide a good description of the observed  emission ($\chidof=400.1/361$ and $\chidof=394.5/360$ for the slab and spherical  geometry respectively) with no clear residuals. In both these scenarios the reflection  fraction is similar to the one measured with the \textsc{pexrav} model  ($R> 1.6$ in both cases).

The amount of reflection is consistent with  the \sax\
measurements  of \sorg\ (\citealt{Risa2002,Dadina2007}), for which the authors report a
reflection fraction ranging from 0.7 to 2.0 (\citealt{Risa2002}), while it is
remarkably higher than the value reported from a {\it RXTE} measurement ($R=0.4\pm
0.1$, \citealt{Rivers}).  The  apparent discrepancy  between  these
measurements could be ascribed to the combination of different effects; among
them, variability of the primary continuum and of the amount of absorption (see \S
4). In particular, the \suzaku\ measurement appears to be, at a first glance, consistent
with the scenario proposed for the \sax\ observations, where the increase
of the reflection fraction was ascribed to a   Compton-reflected component
remaining constant despite a drop  in the primary power-law flux.  
We note that the  {\it RXTE} observations were performed in two campaigns one in 1996 and one in 2003, with  94\% of the total good exposure time being  from the 1996 campaign. During these observations the observed  2--10 keV was  a factor of two higher than during the \suzaku\ one,  and thus the lower reflection fraction could be in agreement with this scenario.\\

However,
as already suggested by \citet{Rivers}, the scenario could be far more
complex and also indicative   that the reflection component normalisation
is responding to a different past illuminating flux.   We note however that
not only do these works  assume  different values  for the inclination 
angle, which could affect the measurement of the reflection fraction, but that also the model itself which is adopted for the reflected component is not flawless, indeed it assumes that the reflector is a semi-infinite slab  and also its density  is assumed to be infinite. 
More importantly the energy band   and spectral resolution of the observations have a strong impact on the measurement of the continuum parameters; for  example, the {\it RXTE} observations have a lower resolution at the energy  of the iron line with respect to the \sax\  and \suzaku\  ones, and this has a strong impact on the measurement of the   Fe K line/edge properties  as well as on the measurement of the amount of reflection and  the intrinsic  $\Gamma$. 
 The amount of reflection  also strongly depends on the model assumed for the X-ray absorber, indeed including  the effect of the Compton-down scattering would increase the normalisation of the primary emission  and not  of the reflected component and thus lower the value of the reflected fraction. \\

 A more detailed description of the variability
of this source is presented in \S\S 4 and 4.1, where we compare the
historical X-ray spectra obtained for this source, showing that both the amount
of absorption and possibly  the primary continuum are   varying,  while the  reflected component  and the $\nhsym$ of a distant reprocessor remain rather constant.  We present  a summary of the main parameters of the X-ray
emission that can be derived assuming  both the standard models and    the  new  model for a toroidal reprocessor, i.e. the \textsc{mytorus} model \citep{Mytor}.

\section{Evidence for a variable absorber}
\label{section:XMM}

\begin{figure}
\begin{center}
 \resizebox{0.5\textwidth}{!}{
\rotatebox{-90}{
\includegraphics{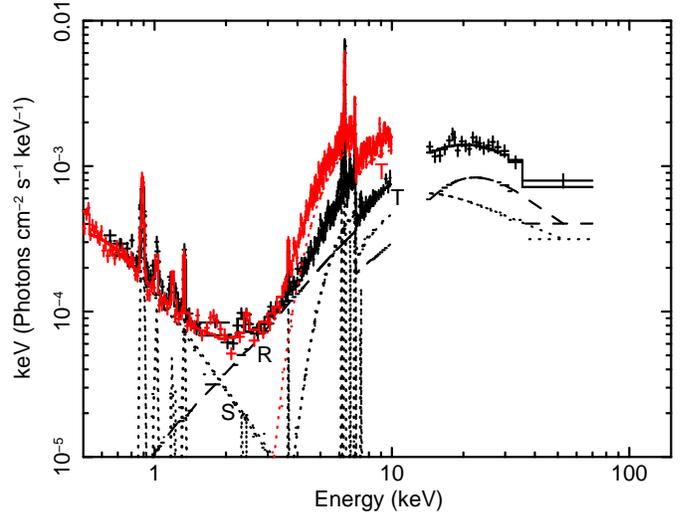}
}}
\caption{Comparison  between the X-ray emission measured with \suzaku\ in 2007  (black data points in the electronic version) and \xmm\  (red data points in the electronic version) in 2001. The continuum model is  a standard model; composed of a primary power-law  component transmitted trough a neutral absorber (labelled with T), a scattered power-law component (labelled with S) and a Compton reflected component (labelled with R and modelled with  \textsc{pexrav}).
For the simultaneous fitting we  allowed to vary both   the $\nhsym$ and the normalisation
of the primary power-law component.  
\label{fig:xmm_vs_suzaku.eps}
}
\end{center}
\end{figure}

 In Fig.~\ref{fig:xmm_vs_suzaku.eps} we compare the \xmm\
 (red) and the  \suzaku\ XIS  \& HXD (black) data; a clear  difference in the curvature is
 present between 4 and 8 keV, which   is most likely  due to a change in the  amount of
absorption of the primary radiation.  \\

To test this hypothesis we applied the \suzaku\ best-fit model to the
\xmm\ spectrum,  allowing     the \feka\ emission line parameters    as well as all the continuum parameters   free to vary  (see Table \ref{tab_continuum}).   
We found a statistically acceptable fit  (\chidof\ = 633.4/503), which  unveiled that the main
difference  between these two observations can be explained with a
lower column density of the  intrinsic absorber ($\nhsym ^{\mathrm {XMM}} =5.0\pm 0.3 \times10^{23}$ \nh and $\nhsym ^{\mathrm {SUZAKU}} \sim 8.2\pm 0.6\times10^{23}$ \nh).   
The normalisation of the primary power-law component also  varied between the two observations as well as the amount of reflection;   however,  we note  that there could be some  degeneracy between the  slope of the
primary  power-law component  and the amount of reflection when lacking a simultaneous high energy
observation.  We note that without allowing the column density to vary,  we could not reproduce the different 2--10 keV spectral curvature observed during the \xmm\ observation.   \\

 To break this degeneracy and better understand the  variability  of \sorg\  we reanalysed the 3 \sax\ observations  of  \sorg\  (hereafter SAX1, SAX2 and SAX3 see Table 1) and, since we are  mainly interested in the hard X-ray emission and variability of the spectral curvature,  we considered  only the MECS and PDS data in the 2--10  keV  and 15--200 keV  energy range respectively. We adopted the best-fit model of the \suzaku\ data and   we  fixed the components responsible for the   soft X-ray emission  to the  \suzaku\ values.  We found  that a simple change of the amount of   X-ray absorption  and the  photon index   cannot explain the observed variability.  We   then  allowed also  the nomalization of the primary continuum and the \feka\  line parameters to vary while constraining the normalization of the reflected component to scale with the primary continuum (i.e. we fixed the ratio $R$,  between the primary and reflected component to the one measured during the \suzaku\ observation). This model represents a situation where  the  reprocessor responsible for the reflected component responds to the variability of the primary continuum and thus it  implies that this absorber should be close to the primary X-ray source.  This model  did not provide a good fit to  SAX1, SAX2 observation (\chidof\ =285.9/145,  \chidof\ = 159.7/108), while it is  a statistically acceptable fit for the last \sax\ observation (\chidof\ =110.9/100), during which \sorg\ was in a state similar to the \suzaku\ observation.  \\

Upon allowing also  the ratio between the  intensity of the reflection  component  and the primary continuum free to vary  (i.e. the parameter $R$)  the fits were acceptable and  we found  \chidof\ =159.6/144,  \chidof\ = 101.5/107  and \chidof\ = 108.6/99  for the SAX1, SAX2 and SAX3 observations, respectively.  The parameters of these best-fits are reported in Table \ref{tab_continuum};  we note that they are in agreement with the results  previously presented in \citet{Matt04}, \citet{Risa2002} and \citet{Dadina2007}. We found that the Fe K line emission complex is rather constant and also note there is no evidence  for variability of the intensity of the reflection component with the \sax\ observations.

 \subsection{A more physical model}
 
 This simple test, as described above,  shows us that the variability properties of \sorg\ are more complex than a simple variation of the amount of absorption. We also note that the intrinsic photon index as well as the Fe K emission line complex are not  variable, while variations are present in the column density and 
   intensity of the  continuum level   and  thus in the ratio of the reflection component versus the primary continuum. However the absolute flux of the reflection component is consistent with being constant. This in turn tells us  that there is a  rather stable reprocessor, which is responsible for the Fe K emission lines and the reflected component and   which  does not  appear to respond to the variability of the primary continuum.  This could be indicative of   distant reprocessor,  which  does not  respond to the variability of the primary continuum. Alternatively as we will discuss later this could indicate  a clumpy absorber where  the overall distribution of clouds remains rather constant. Given the limitations  of the  {\sc{pexrav}} model, already outlined above (i.e. the geometry and density assumed for the reflector),  this simple model   does not allow us to derive strong constraints on the  true nature of the absorber.  Furthermore,   by adopting  this non physical model  the   temporal properties  of the reprocessor   (i.e. the variability of the amount of  line-of-sight 
absorption  and Compton reflection) could be   highly uncertain  and degenerate with respect to the variability of the  primary continuum.  Finally,  the column densities of the X-ray absorber  are in the range where the correction for the Compton-down scattering starts to be important.
 Thus  we  first included an  additional absorber  (\textsc{CABS} model in \textsc{XSPEC}) to account for the effect of the Compton-down scattering. We found a similar trend     (as the one reported in Table~4)  in the normalisations of the primary power-law components and thus in  the intrinsic 2--10 keV luminosities, albeit with a larger spread.  \\

 Therefore we decided  to reanalyse the available spectra using the  most recent model for the toroidal  reprocessor \citep{Mytor}, which    correctly accounts for the emission  expected in transmission  (hereafter zeroth order continuum) and reflection  and  also includes  the expected Fe K emission lines (\feka, \fekb\ and the Compton shoulder).
The calculations at the basis of this new model  are all fully   relativistic and  valid   for $\nhsym$ in  the Compton-thin and  Compton-thick regimes. 
This new  model  assumes a uniform and essentially neutral  toroidal reprocessor  with an opening angle of $60 ^{\circ}$ with respect to the axis of the system,     while we note that the \textsc{pexrav} model assumes a  disc/slab geometry for the reflector and thus the  parameters derived from  this model, such as the covering factor,  can not be directly related to a covering factor of the putative torus as well as  to the  line-of-sight  column density.   In summary in the \textsc{mytorus} model all the different continuum components (reflected and transmitted  components) and the fluorescent emission lines are all treated self-consistently and can thus   be all directly related to the key parameters of the matter from which they originate. 
By adopting this model we were able  to   determine which   component  dominates in each energy band  (reflected
or transmitted components),  assess their variability properties  and   thus better understand the global
distribution of the absorber. \\

\subsubsection{ A new implementation of the Mytorus  model applied to the \suzaku\ observation}
The  standard  \textsc{mytorus} model, developed for \textsc{xspec}, is composed of   three tables of reprocessed spectra calculated  assuming that the input spectrum is a power law. These tables correspond to the main model components  expected from the interaction of the primary power-law component with a reprocessor that has a toroidal geometry:  the  distortion to the zeroth-order (transmitted) continuum   (MYtorusZ), the reflected continuum (MYtorusS), and the \feka, \fekb\ emission-line spectrum   (MYtorusL).  MYtorusZ  is a multiplicative table that  contains the  pre-calculated   transmission factors   that distort the incident continuum at all energies  due  photoelectric absorption.\\ 

We  first applied this toroidal-reprocessor model ({\sc mytorus}  \citealt{Mytor}) to the \suzaku\ observation. 
The model setup is the following: \\

\textsc{phabs$\times$ (apec  + apec + A$_{\rm soft}\times$zpowerlw  + 6 GA$_\mathrm{em}$ +
\mbox{\ion{Fe}{xxv}}  +  Ni K$\alpha$ +  MYtorusZ$\times$zpowerlw +    A${\rm_R}\times$ MYtorusS  + A$\rm{_L}\times$ gsmooth
$\times$ MYtorusL)}\\

We also included the \fexxv\ and \nika\ emission
lines as well as a soft  power-law component  ($A_{\rm soft}\times$zpowerlw) that represents  scattering off optically-thin ionised   gas (warm or hot), which are not included in the  \textsc{mytorus} model. For the soft X-ray emission we kept the  6 Gaussian emission lines  (\textsc{GA$_{\rm{em}}$}; see Table~2) and we also included two thermal emission components, which allowed us to tie the photon index
of the soft power-law component to the primary one as expected from scattering off optically-thin  ionised  gas. The  normalisations of the   \textsc{mytorus} components and of the  soft  power-law component are all tied together, while the value of the relative normalisations are included in the factors   $A_{\rm soft}$, $A{\rm_R}$ and   $A{\rm_L}$, which are in turn the relative normalisations of the soft power-law component,  of the reflected component and of the emission lines.  Since we expect the size scale of the scattering/reflecting and line emitting regions to be similar, we  initially   set $A{\rm_R} = A{\rm_L}=1$  to be equal and we set them to  1. We note that we can not interpret  any  difference  between these two factors  as a difference in the size scales of these zones since   these  constants also include  the effects of the transfer functions of the reflected continuum and line spectrum.    The     \textit {gsmooth} component is the broadening of the Fe  K emission lines  and it is actually composed of two  Gaussian convolution components, one is the actual broadening of the \feka\ emission line  while the second component accounts for the weak residual instrumental broadening as measured with the calibration sources ($\sigma< 15 $ eV, with a $\sqrt{E}$ dependence). \\
  
  \begin{figure}
\begin{center}
 \resizebox{0.5\textwidth}{!}{
\rotatebox{-90}{
\includegraphics{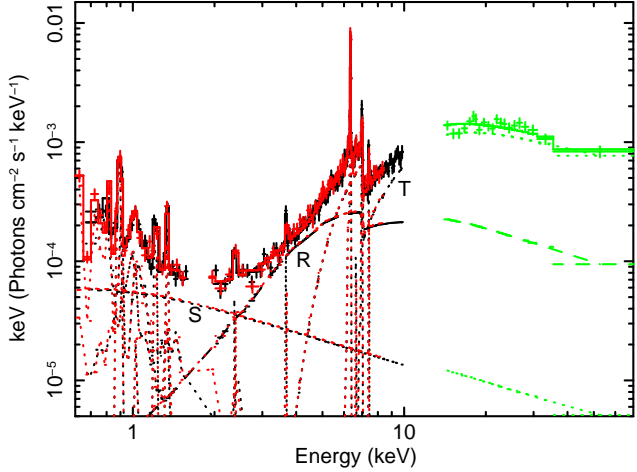}
}}
\caption{The 0.5--70 keV  \suzaku\  spectra when fitted with the \textsc{mytorus} model (in the electronic version: XIS-FI, black; XIS1, red; HXD-PIN  green). The dashed lines represent the reflected  component (marked with R), the dotted lines are the  transmitted component (marked with T) and the scattered component (marked with S) from  ionised gas.
\label{fig:tahir.ps}
}
\end{center}
\end{figure}

An inspection of the \suzaku\ spectra  shows that above 8 keV we are dominated  by the  primary  component, transmitted   through the reprocessor, which is mainly constrained by the high-energy excess above 10 keV, while the reflected component dominates below 8 keV, where the strong emission lines from the Fe K$\alpha$, Ca K$\alpha$ and {S\,\textsc{xiv}}  K$\alpha$  are  present (see Fig.~ 7).  This is analogous to what is observed adopting the old {\sc{pexrav}} model (see Fig.~\ref{fig:model1_data.ps}, upper panel) and it is also in agreement with the variability of the normalisation of the primary power-law component as suggested in section \S4. \\

We  note that in the standard configuration of the {\sc mytorus} model,  we cannot account self-consistently  for the  requirement of a strong transmitted component emerging at higher energies, the  intense Fe K emission lines and a dominant reflected component below 8 keV. 
In particular,  if we adopt the standard toroidal geometry with  the inclination angles (between the axis of the torus and the observer's line of sight) of the two components tied together   and if we  do not   allow  the
normalisations of the  reflected component   to vary  with respect to the  normalisation of the   zeroth order    continuum strong residuals are present below 10 keV. Keeping   $A{\rm_R}$ and   $A{\rm_L}$   tied to each other, we found that a cross-normalization factor  of  $A{\rm_R}\sim 3$  is indeed required to  reproduce the shape of the 4--10 keV continuum and the intensity of the Fe K emission lines, where the reflected  component dominates.  This forces   the inclination angle between the axis of the torus and the observer's line of sight  to a grazing value ($\sim 60^{\circ}$). \\

Although we cannot  rule out   this  scenario, we must allow for the
possibility of a different geometry taking  into account all the information that we  obtained from
the   historical X-ray observations of \sorg\  which suggested that: a) the  column density of  the line of sight  ({\it los})
absorber varies; b)  the flux   of the \feka\ remains rather stable, suggesting the presence of a  constant and  distant reprocessor and c) the primary continuum is
also  variable.  Physically, the situation
we want to  model   corresponds to a
patchy reprocessor in which the reflected continuum is observed
from reflection in matter on the far-side of the X-ray source, without
intercepting any other ``clouds,'' and the zeroth-order continuum
corresponds to extinction by clouds in the line-of-sight. In practice this corresponds  to allowing the
column densities of the zeroth-order and reflected continua to
be independent of each other; we thus  followed the methodology discussed by \citealt{Yaqoob4945} applied to the modelling of the broadband X-ray emission of NGC 4945.  By decoupling these two components, we can also allow the reflected and transmitted  components to  have a different  temporal behaviour. We can do this by decoupling the  inclination angle
parameters for the line-of-sight (zeroth-order) continuum
passing through  the reprocessor and for  the reflected continuum
from the reprocessor and allowing the column densities   responsible for the reprocessing of the primary emission to  be independent. The reflected continuum (and the
fluorescent line emission, which is tied to it) is not extinguished by another column.  \\

The inclination
angle of the zeroth-order component is now irrelevant (so it is
fixed at 90 degrees), and the inclination angle for the
reflected continuum is fixed at 0 degrees because the effect of the inclination angle
on the shape of the reflected continuum is not sufficiently large (in terms of spectral fitting) if the
reflected continuum is observed in reflection only.  Furthermore,   since we are  trying to model a patchy reprocessor as suggested from the column density variations,  
the inclination angle  may not be meaningful. We  note also that we cannot  interpret the  ratio between the  normalisations of the reflected  and transmitted  components  simply as a  covering factor of the reprocessors. Although the constant in  front of the reflected continuum 
($A_{\rm R}$) does contain some information on the covering
factor, that information cannot be decoupled from the
effect of time delays between variability of the direct X-ray
continuum and the reprocessed  X-ray continuum. This is
because  the light-crossing time of the reprocessor is likely to be much
longer than the direct X-ray continuum variability timescale, so
the magnitude of the reflected  continuum corresponds
to the reprocessed direct continuum that is averaged over a timescale
that is longer than the reprocessor light-crossing time. This decoupling of the    \textsc{mytorus} model is  close  to the standard procedure  used while fitting with {\textsc PEXRAV} plus an absorbed power-law component. However there are several  differences, in particular the column density of the reflector is also a free parameter and the Fe emission line intensities are calculated self consistently.\\

 The model then yields a \chidof\ = 392.5/359   and a  mean line-of-sight column density  of
$\nhsym=9.4$\errUD{0.2}{0.2}$\times 10^{23}$ \nh , while the angle-averaged column density of the reflector  (out of the
line-of-sight) is $\nhsym=2.6$\errUD{0.2}{0.2}$\times 10^{23}$ \nh, where also the Fe K emission lines are produced.   The photon index is found to be
$\Gamma=1.68$\errUD{0.03}{0.03} and the normalisation of the  primary continuum is $1.91\errUD{0.16}{0.25}\times10^{-2}$ ph cm$^{-2}$ s$^{-1}$.
 We also allowed  the constant for the  normalisation  of the
reflected  continuum ($A_{\rm R}$)   to vary and we found that it is consistent with 1 ($A_{\rm R}=1.1\pm 0.2$). Finally, we note that now    the measured  velocity broadening of the \feka\  emission line is  $\sigma_v<29$ eV.

 \begin{table*}
\caption{Summary of the X-ray emission lines detected in the 6--8 keV energy range. The energies of the lines are   quoted in the rest
frame. Fluxes and identifications are reported in column  2 and 3.    The EW are reported in column 4 and they are 
calculated against the total  observed continuum at their respective energies. In column 5 the
improvement of fit is shown with respect to the continuum model, the value for the model with no lines 
is \chidof\ =1918.9/375.  \label{table:Fe_lines} }
\begin{tabular}{ccccc}
\hline
Energy      & Flux  &ID        & EW & $\Delta \chi^2$  \\
 (keV)       & ($10^{-6}$ph cm$^{-2}$ s$^{-1}$) & &(eV) &\\
 
    (1)   &  (2)   &  (3)    &  (4)   & (5)   \\

\hline
     &     &      &         &  \\
   6.408\errUD {0.005}{0.004}& 52.4\errUD{3.7}{5.1} & \feka\  & 490\errUD {40}{50}     & 1427.4   \\   
 &   &   &     &\\

 7.07\errUD {0.02} {0.02}& 10.0\errUD {1.8}{1.8}  &  \fekb\ & 81\errUD  {23}{13} & 82.0\\
  &   &   &      &\\
 
 6.73  \errUD {0.05}{0.06} & 3.1\errUD {1.6}{1.6} &  \fexxv&   27 \errUD {15}{13} & 8.1\\
   &   &   &       &\\
 7.50\errUD {0.07}{0.08}&  2.3\errUD {1.5} {1.4}& \nika &37\errUD{24}{23}  & 6.7\\
    &   &   &     & \\
 \hline
\end{tabular}
\end{table*}

\begin{table*}  
 \caption{Comparison between the  best fit values for the  continuum and \feka\ emission line     
for the \suzaku\, \xmm\ and \sax\  observations.  The fluxes   are corrected only for Galactic absorption.} 
\label{tab_continuum} 
\begin{tabular}{lcc  c cc} 
\hline
    Parameter  &  \suzaku &\xmm\ &  SAX1 & SAX2 &SAX3 \\ 
 \hline
& &  \\\
DATE &2007-12&2001-01&1997-07&1998-07&1999-01\\
 $N_{\rm H}$ ($10^{23}$ \nh) & $8.2_{-0.6}^{+0.6}$  & $5.0^{+0.2}_{-0.3} $& 7.0\errUD{0.4}{0.5}& 6.2\errUD{0.7}{0.3}&7.2\errUD{2.9}{1.6}\\
 
 $\Gamma_\mathrm{Hard}$ &$1.81_{-0.04}^{+0.04}$ & $1.76_{-0.04}^{+0.05}$ &1.77\errUD{0.03}{0.05} & 1.72\errUD{0.07}{0.1}&1.6\errUD{0.1}{0.1}\\

 Normalisation ($10^{-2}$ ph cm$^{-2}$ s${-1}$) & $0.64_{-0.11}^{+0.13}$& $ 1.25_{-0.06}^{+0.06}$ & 2.52\errUD{0.22}{0.33} &1.70\errUD{0.19}{0.23} &0.62\errUD{0.48}{0.23}\\

$\Gamma_\mathrm{Soft}$ & 3.8\errUD{0.2}{0.2}& 3.7\errUD{0.1}{0.1}&.. &..&..\\ 

 Normalisation ($10^{-4}$ ph cm$^{-2}$ s$^{-1}$) & $1.70_{-0.09}^{+0.09}$& $1.37_{-0.05}^{+0.06}$& ..&..&..\\

 $A_\mathrm{pexrav}$  ($10^{-2 }$ph cm$^{-2}$ s$^{-1}$)   &$ 1.19^{+0.09}_{-0.09 }$ &  $0.88^{+0.14}_{-0.13 }$& 1.6\errUD{0.2}{0.3}&1.2\errUD{0.3}{0.3}&1.0\errUD{0.5}{0.4}\\

Fe K$\alpha$   (keV)       & $6.408 _{-0.004} ^{+0.005}$  &$6.39 _{-0.02} ^{+0.04}$  &6.39\errUD{0.10}{0.09} & 6.58\errUD{0.11}{0.12}  &6.42\errUD{0.07}{0.07}  \\

 $I_\mathrm{FE}$  ($10^{-5 }$ph cm$^{-2}$ s$^{-1}$)    &$ 5.24^{+0.51}_{-0.37}$ &  $4.36^{+0.43 }_{-0.41}$ & 4.7\errUD{1.7}{1.8}&6.7\errUD{2.0}{2.2}& 5.9\errUD{1.5}{1.5}\\

$ EW_\mathrm{FE}$ (eV) & 490\errUD {40}{50}& 190\errUD{20}{100} &140\errUD{80}{60}& 225\errUD{65}{90}&400\errUD{90}{120} \\

F $_{(0.5-2\;\mathrm {keV})}$ ($10^{-13}$ \flux)& $\sim 4.8$  & $\sim  4.9 $ &... &...&... \\

F$_{(2-10\;\mathrm {keV})}$ ($10^{-11}$  \flux ) & $\sim 0.6$ & $\sim 1.2 $ & $\sim 1.8$ & $\sim 1.6$ & $\sim 0.8$\\

L$_{(2-10\;\mathrm {keV})}$ ($10^{43}$  \lum ) & $\sim 0.9$ & $\sim 1.5 $ & $\sim 2.8$ & $\sim 2.2$ & $\sim 1.0$\\
  \hline
\end{tabular} 
 \end{table*}

\begin{table*}  
 \caption{Comparison between the  best fit values for the  continuum and $\nhsym$     for the \suzaku, \xmm\  and \sax\ observations when fitted with the  \textsc{mytorus} model.  The fluxes are corrected  only for Galactic absorption,   while the luminosities are corrected also for the intrinsic absorption, which  includes the effect of the Compton-down scattering.} 
\label{tab_mytor} 
\begin{tabular}{lcc  c cc} 
\hline
    Parameter  &  \suzaku &\xmm\ &  SAX1 & SAX2 &SAX3 \\ 
  &2007-12&2001-01&1997-07&1998-07&1999-01\\
 \hline
& &  \\\

  $\Gamma$ &$1.68_{-0.03}^{+0.03}$ & $1.68^{f}$ &1.62\errUD{0.03}{0.03} & 1.66\errUD{0.04}{0.05}&1.63\errUD{0.08}{0.07}\\
 Normalisation ($10^{-2}$ ph cm$^{-2}$ s${-1}$) & $1.91_{-0.25}^{+0.16}$& $ 1.56_{-0.12}^{+0.05}$ & 2.89\errUD{0.32}{0.35} &2.98\errUD{0.49}{0.52} &2.12\errUD{0.87}{0.61}\\

 $N_{\rm H\; transmitted}$ ($10^{23}$ \nh) & $9.38_{-0.24}^{+0.25}$  & $4.66^{+0.10}_{-0.27} $& 6.33\errUD{0.29}{0.35}& 6.46\errUD{0.45}{0.46}&8.58\errUD{0.10}{0.96}\\
 $N_{\rm H\; reflected}$ ($10^{23}$ \nh) & $2.56_{-0.17}^{+0.20}$  & $3.54^{+1.26}_{-0.41} $& 2.64\errUD{0.88}{0.50}& 2.40\errUD{0.85}{0.89}&2.37\errUD{1.29}{1.05}\\
F$_{(2-10\;\mathrm {keV})}$ ($10^{-11}$  \flux ) & $\sim 0.6$ & $\sim 1.2 $ & $\sim 1.7$ & $\sim 1.6$ & $\sim 0.9$\\

L$_{(2-10\;\mathrm {keV})}$ ($10^{43}$  \lum ) & $\sim 2.5$ & $\sim 2.1 $ & $\sim 4.0$ & $\sim4.1$ & $\sim 3.0$\\

  \hline
\end{tabular} 
 \end{table*}

 \subsubsection{Mytorus  model for \suzaku, XMM-Newton and \sax}

We then applied the same model to the \xmm\ observation. For simplicity, since there is no evidence of variability of the soft X-ray emission and taking into
account that the Gaussian emission lines plus the thermal components are  simple phenomenological models,  we
decided to keep  fixed the main parameters of the latter to the \suzaku\ best-fit model.  Furthermore, since we lack of  simultaneous observation above 10 keV we also fixed the photon index to the  one measured with \suzaku\ and  for simplicity at first we kept the constant of the relative emission line component ($A_{\mathrm L}$) fixed to 1.
We found that the out of {\it los }column density
   was comparable to the one measured during the \suzaku\ observation
($\nhsym=3.5$  \errUD{1.3}{0.4}$\times 10^{23}$ \nh)  while the {\it los }absorbing column
density  was $N_{\rm {H}}= 4.7^{+0.1}_{-0.3} \times 10^{23}$ \nh ($\chi^2/$dof=$600.5/498$).
 
 In contrast
with the previous modelling with \textsc{pexrav}, we can now attempt to investigate also the relative intensity of
the zeroth-order and reflected components. We found that the intensity of the zeroth order changed from
$(1.91\errUD{0.16}{0.25})\times10^{-2}$ ph cm$^{-2}$ s$^{-1}$  to $(1.56\errUD{0.05}{0.12})\times10^{-2}$ ph cm$^{-2}$ s$^{-1}$
during the \xmm\  and \suzaku \ pointing respectively.This suggests that  no strong variation of the primary continuum is required to explain  the observed 2--10 keV spectral differences, but the main driver  of the variations is the change in the column density of the line-of-sight absorber.\\

Finally, we applied the same model to the 3 \sax\ observations, allowing also   the photon index to vary, and we found that the column density of the out of the {\it los} absorber remained stable and it was comparable   (within the errors) to the one measured with \suzaku\ and \xmm,   ($\nhsym=2.6$  \errUD{0.9}{0.5}$\times 10^{23}$ \nh,  $\nhsym=2.4$  \errUD{0.9}{0.9}$\times 10^{23}$ \nh\ and $\nhsym=2.4$  \errUD{1.3}{0.1}$\times 10^{23}$ \nh\ for SAX1, SAX2 and SAX3 respectively). The  column density of the {\it los} absorber varied with a similar trend as the one measured with the \textsc{pexrav}-based model ($\nhsym=6.3$  \errUD{0.3}{0.4}$\times 10^{23}$ \nh,  $\nhsym=6.5$  \errUD{0.5}{0.5}$\times 10^{23}$ \nh and $\nhsym=8.6$  \errUD{0.1}{1.0}$\times 10^{23}$ \nh for SAX1, SAX2 and SAX3 respectively), the intensity of the primary continuum   components also  varied.  In particular, the intensity of the primary continuum was higher in the  SAX1  and SAX2 observation and  in  SAX3.  However, due to the lower statistics of the \sax\ data these measurements have a large error which prevent us  from  deriving  a clear picture.   We note   that there is no evidence for a  variation of the photon index.

\section{Discussion and conclusions}

Detailed  X-ray spectral analysis of the \suzaku\ data  confirms the complexity of the X-ray emission from \sorg. Thanks to the wide-band spectrum covering from 0.6 keV to 70 keV, we have now obtained the  most reliable  deconvolution of all the spectral components. The X-ray continuum is composed of three components: a heavily absorbed power-law component, a reflected component and a  weak soft scattered component.    We  analysed the \suzaku\ and the  historical observations of \sorg\ adopting both a standard model as well as a new toroidal reprocessor model. 
With either of these two approaches we  found that during the \suzaku\ observation the 2--10 keV emission of \sorg\ was dominated by  the reflected emission, while above 10 keV the spectrum was dominated by the highly absorbed  transmitted component.\\

The soft X-ray emission can be well described with a superposition of a power-law component, which is considered to be the scattered light from   ionised, optically-thin gas, and several emission lines. These emission lines were already detected with the ASCA observation (\citealt{Comastri98}). As already shown  with the \xmm\ observation \citep{Matt04}, the  wide range of ionisation  implied by these lines is indicative of the presence of at least two  photo- or collisionally- ionised  emitters. This is confirmed by a recent deep \xmm\ observation obtained by our group within a monitoring program of \sorg.  The analysis of the single observations showed  also that as seen in other Seyfert 2s the soft X-ray emission did not vary, implying that the emitters responsible for this emission are located  outside the variable  X-ray absorber (\citealt{Marinucci4507}, Wang et al  private communication).\\

 \subsection{The X-ray absorber}
In the last two decades NGC~4507, which is one of the X-ray brightest and nearby Seyfert 2 galaxies, has been observed several times with all the different X-ray observatories; NGC~4507 displayed an observed  2--10 keV flux ranging from $0.6-1.3\times10^{-11}$\flux; furthermore these X-ray observations showed long-term $N_{\mathrm{H}}$ variability, which changes by a factor of 2, and also  possible variability of the intrinsic continuum (\citealt{Risa2002}).   The \suzaku\ observation   caught the source with a   low  observed 2--10 keV flux, similar to the last \sax\ observation (SAX3), which was also characterised by the highest measured column density of the X-ray absorber ($ 8-9 \times 10^{23}$ \nh).  Our analysis of the  \xmm\, \sax, and \suzaku\ observations shows that, independently from the assumed model (i.e.  \textsc{mytorus} or the standard  \textsc{pexrav} one)  the X-ray absorber in \sorg\ varies from  $\sim 5\times 10^{23}$ \nh\ to $\sim 9\times 10^{23}$ \nh\  on $\sim 6$ months time scale between the observations. \\

Assuming  a spherical geometry for the
obscuring clouds, and that they are moving with Keplerian velocities (as in the case of  NGC~1365; \citealt{Risaliti07}), then there is a first order simple relation between  the crossing time of such obscuring cloud, the  linear dimension and distance of the obscuring cloud  and  the size of  the X-ray source  (see \citealt{Marinucci4507}).  From  this work, since  no variability is found on a short time scales, during the long \suzaku\ observation, we can infer that the variability occurs on a time scale between 2 days (as the elapsed time of the \suzaku\ observation was $\sim 180$ ks) and six months (as the elapsed time between the  second and the third \sax\ observations).  Following   the same argument proposed for NGC~1365 (\citealt{Risaliti07}), where similar assumptions are used, then  a possible scenario  could be  that the variable absorber is located a distance greater  than 0.01 pc  from the  X-ray source (i.e.  not   closer than the Broad Line Region).  
Thus a possible scenario, where the classical uniform absorber still exists is that there are multiple absorbers. One  is the classical  and uniform  pc-scale absorber (i.e. the torus), which is responsible for the Fe K emission line and the constant reflected component; while  the $N_{\mathrm{H}}$ variability requires the presence of a second and clumpy absorber that  could be coincident with the outer BLRs. We note that the observations presented here do not span  the possible time-scale  expected for a variable absorber located  between the BLR and the pc-scale absorber.   \\

However a simpler    scenario could be  that the variability is due to a certain degree of clumpiness of a single absorber itself,  as for the  clumpy torus  model (\citealt{Nenkova2002,Nenkova2008, Elitzur2006}). In this scenario,   if the distribution of clouds remains rather constant,  we can have a  constant reflected component while at the same time  changes  of the line-of-sight $\nhsym$.  Recently our group obtained an \xmm\ monitoring campaign  of \sorg\  consisting of   six observations  performed every 10 days,   this monitoring confirmed that the  $\nhsym$ variability   occurs on relatively long time scale (between 1.5 and 4 months),   supporting the    scenario of a clumpy pc-scale absorber (\citealt{Marinucci4507}). A    long term variability or lack of  certain components cannot be   simply  used to  infer the location of the reprocessor itself in terms of time delays  between the reflected component and the primary continuum.  As discussed in several works on the  spectral variability  of well monitored Seyfert 1s (e.g MCG-6-30-15 \citealt{Miller08}; Mrk~766 \citealt{Miller766}),  the  apparent variability of the observed continuum level  could be due to   changes of the los covering  fraction and     not of the  intrinsic continuum level. Thus the reflected  emission  would remain rather constant  even for closer in and   variable absorbers. \\

  Stronger evidence for the presence of a  pc-scale reflector can be derived from the analysis of the \feka\ emission line profile.  We note that  \feka\ emission line is  rather constant and  narrow with no evidence of an additional  strong  broad component;  its width,  measured with \suzaku,   is $\sigma< 30 $ eV (or $\sigma_v<1400$ km s$^{-1}$)  and consistent  with the  \chandra\ upper limits, $\sigma_v<1200$ km s$^{-1}$ ($FWHM<3000$ km s$^{-1}$). The measured $FWHM$ of the  H$\alpha$ ($\sim 5000$ km s$^{-1}$; \citealt{Moran}) is marginally higher than  the  \chandra\ upper limit on the  \feka\  $FWHM$. This  suggests that the \feka\    is produced  either in the   outer part of the BLRs or  in  a pc-scale absorber; in agreement with  a scenario  where there  is a distant and stable reprocessor, which   could be identified with the classical torus.   A similar  result has been  presented by \citet{Shu2011}, where the authors compared the FWHM of  the \feka\ emission line  (measured with the \chandra\ HETG) of a sample of Seyfert 2s (including \sorg) with the $FWHM$ of the  optical lines. They  suggested that  the \feka\ emitter is a factor 0.7--11 times larger than the optical line-emitting region and located at a distance of about $r\sim 3\times10^4r_g$ (where $r_g$ is the  gravitational radius defined as $r_g=GM/c^2$). In particular from  the estimated  black hole mass  for \sorg\    of M$_{\mathrm{BH}}\sim  4\times10 ^{7}$ M$_\odot$ (\citealt{Bian2007}) and assuming Keplerian motion, the upper limit on its $FWHM$ implies that the Fe K$\alpha$ emission line is produced  at a distance $r>0.02$ pc from the central BH.  We note that assuming the larger BH mass  (M$_{\mathrm{BH}}\sim  2.5\times10 ^{8}$ M$_\odot$) reported by \citet{Winter09}  would place the absorber at a distance $R>0.06$ pc and the location of the \feka\  at $r> 0.1$~pc.\\
  
   In terms of the  global  picture for the  location and structure of the X-ray absorbers, we have now  several  examples of obscured AGNs with short-term variation unveiling that  a significant fraction of the absorbing clouds are located within the BLR. However there is also evidence for the presence of a pc-scale absorber as predicted in the Unified Model of AGNs. This absorber is confirmed by the ubiquitous presence of the narrow \feka\ emission line and the Compton reflection component, which  do not show  strong variability between observations   even   in case of a variable intrinsic continuum and/or variable neutral absorber  (e.g. NGC~7582 \citealt{Piconcelli2007,Bianchi09}, NGC~4945; \citealt{Marinucci4945,Yaqoob4945,itoh}). Another  piece of evidence for the presence of a distant reprocessor comes from the comparison between the  \chandra, \xmm\  and \suzaku\ observations of  the  bright  Seyfert 2 NGC~4945 (\citealt{Marinucci4945,Yaqoob4945}), where  a detailed spectral, variability  and  imaging analysis unveiled that the  emitting region responsible for the \feka\  line and  the  Compton-scattered continuum has a low covering factor and it is most  likely located at a distance $>30-50$ pc.  In this framework the relatively long-term absorption variability shown by \sorg\ confirms  that  the  location and structure of the X-ray absorber is complex  and  that absorption in type 2 AGNs   could occur   on  different  scales and that   there may not be a universal  single and   uniform absorber.\\

Finally, by adopting the standard \textsc{pexrav} model,  even with the broad band X-ray observations available for \sorg, we cannot assess the role of the possible variability of the primary continuum or  estimate the column density of the  reprocessor responsible for the Fe K$\alpha$ emission  line. 
Interestingly,   the fit with the decoupled  \textsc{mytorus}  model, which can mimic either a patchy toroidal reprocessor  or a situation where there are  two reprocessors (one seen in "transmission", dominating the high-energy spectrum, and one  seen in reflection),    allows us to measure these column densities and the possible  variability of the primary continuum. Although  the  decoupled  \textsc{mytorus} model   closely resembles the classical  combination of \textsc{pexrav} (slab reflection component) and an absorbed power-law component,  the  column densities of both the  reprocessors (``reflector'' and ``absorber'') are  treated independently and  self-consistently with the emission of the Fe  K line.
 Table~\ref{tab_mytor} shows that the  line-of-sight obscuration of the reprocessor seen in transmission varies by $\Delta N_{\mathrm{H}}\sim 5\times 10^{23}$ \nh, while there is a constant reprocessor with a column density of $\sim 2-3\times 10^{23}$\nh, which is the one responsible for the Fe K$\alpha$ emission line.  The intrinsic X-ray luminosity ranges from $L_{(2-10\;\mathrm {keV})} \sim 2.1 \times 10 ^{43}$ \lum to  $L_{(2-10\;\mathrm {keV})}\sim 4.1  \times 10^{43}$ \lum.   While we observe intrinsic variation of  primary power-law intensity,  the $\nhsym$ variablity drives the spectral changes between 2-10 keV. Conversely, the reflection and emission line components are not observed to vary.\\

   We note that the   behaviour of the  line-of-sight column and reflection fraction  with respect to the    intrinsic continuum going from the \sax\ to the \suzaku\ observation strongly  depends on the adopted model  for the reprocessor. Indeed  as can be seen  comparing Table~\ref{tab_continuum} and ~\ref{tab_mytor}, by adopting the  combination of \textsc{pexrav}  and an absorbed power-law component, we would infer that \sorg\  was intrinsically brighter and less obscured during the \xmm\ observation. However,   no  such  trend is inferred with the   \textsc{mytorus}  model, where the opposite is the case (i.e. the more absorbed \suzaku\ observation has the higher primary power-law normalisation).
 Only future monitoring campaigns   with broad band observatories such as ASTRO-H  (i.e. with instrument with an high effective area above 10 keV as well as high spectral resolution at the \feka\ line)  will allow   monitoring  of  sources like \sorg. These observations will allow  to   investigate the variability of the harder continuum  simultaneously providing a detailed investigation of the profile of the \feka\ emission line,  thus  establishing  the geometry  and location of the ``stable''  and variable reprocessors.  \\

  \section*{ACKNOWLEDGMENTS} 
  This research has made use of the NASA/IPAC Extragalactic Database (NED) which is operated by the Jet Propulsion
Laboratory, California Institute of Technology, under contract with the National Aeronautics and Space
Administration. 
We  thank the referee for her/his suggestions that  improved this paper.

 \end{document}

%% file: refs.tex
\def\araa{ARA\&A}%
\def\apj{ApJ}%
\def\apjl{ApJ}%
\def\apjs{ApJS}%
\def\aap{A\&A}%
\def\aapr{A\&A~Rev.}%
\def\aaps{A\&AS}%
\def\mnras{MNRAS}%
\def\pasp{PASP}%
\def\pasj{PASJ}%
\def\nat{Nature}%
\def\physrep{Phys.~Rep.}%

%% file: ms_vbraito_3.bbl
\begin{thebibliography}{99}
\bibitem[Antonucci (1993)]{Antonucci} Antonucci, R.\ 1993, \araa, 31, 473 
\bibitem[Arnaud(1996)]{xspecref} Arnaud, K.~A.\ 1996, Astronomical Data Analysis Software and Systems V, 101, 17 
\bibitem[Awaki et al.(1991)]{Awaki1991} Awaki, H., Kunieda, H., 
Tawara, Y., \& Koyama, K.\ 1991, \pasj, 43, L37 
\bibitem[Bassani et al.(1995)]{Bassani1995} Bassani, L., Malaguti, 
G., Jourdain, E., Roques, J.~P., \& Johnson, W.~N.\ 1995, \apjl, 444, L73 
\bibitem[Bassani et al.(2006)]{Bassani2006} Bassani, L., Molina, 
M., Malizia, A., et al.\ 2006, \apjl, 636, L65 
\bibitem[Beckmann et al.(2009)]{Beckmann2009} Beckmann, V., Soldi, S., Ricci, C., et al.\ 2009, \aap, 505, 417 
\bibitem[Behar et al.(2010)]{Behar} Behar, E., Kaspi, S., Reeves, J., et al.\ 2010, \apj, 712, 26 
\bibitem[Bian \& Gu(2007)]{Bian2007} Bian, W., \& Gu, Q.\ 2007, \apj, 657, 159 
\bibitem[Bianchi et al.(2006)]{Bianchi06} Bianchi, S., Guainazzi, M., \& Chiaberge, M.\ 2006, \aap, 448, 499 
\bibitem[Bianchi et al.(2009)]{Bianchi09} Bianchi, S., Piconcelli, E., Chiaberge, M., et al.\ 2009, \apj, 695, 781 
\bibitem[Bianchi et al.(2012)]{Bianchi2012} Bianchi, S., Maiolino,  R., \& Risaliti, G.\ 2012, Advances in Astronomy, 2012,  
\bibitem[Boldt(1987)]{Boldt} Boldt, E.\ 1987, \physrep, 146, 215 
\bibitem[Comastri et al.(1998)]{Comastri98} Comastri, A., Vignali, C., Cappi, M., Matt, G., Audano, R., Awaki, H., \& Ueno, S.\ 1998, \mnras, 295, 443 
\bibitem[Dadina(2007)]{Dadina2007} Dadina, M.\ 2007, \aap, 461, 1209
\bibitem[Dickey \& Lockman(1990)]{Dickey} Dickey, J.~M., \& Lockman, F.~J.\ 1990, \araa, 28, 215 
\bibitem[Elvis(2000)]{Elvis2000} Elvis, M.\ 2000, \apj, 545, 63 
\bibitem[Elvis et al. (2004)]{Elvis04}Elvis, M., Risaliti, G., Nicastro, F., Miller, J. M., Fiore, F., \& Puccetti, S. 2004, ApJ, 615, L25
\bibitem[Elitzur \& Shlosman(2006)]{Elitzur2006} Elitzur, M., \& Shlosman, I.\ 2006, \apjl, 648, L101 
\bibitem[Elitzur(2012)]{Elitzur2012} Elitzur, M.\ 2012, \apjl, 747, L33 
\bibitem[Fukazawa et al.(2011)]{Fukazawa} Fukazawa, Y., Hiragi, K., Mizuno, M., et al.\ 2011, \apj, 727, 19 
\bibitem[Gruber et al.(1999)]{Gruber} Gruber, D.~E., Matteson, J.~L., Peterson, L.~E., \& Jung, G.~V.\ 1999, \apj, 520, 124 
\bibitem[Guainazzi \& Bianchi(2007)]{Guainazzi07} Guainazzi, M., \& Bianchi, S.\ 2007, \mnras, 374, 1290
\bibitem[Horst et  al.(2006)]{Horst2006} Horst, H., Smette, A., Gandhi, P., \& Duschl, W.~J.\ 2006, \aap, 457, L17 

\bibitem[Itoh et al.(2008)]{itoh} Itoh, T., Done, C., Makishima, K., et al.\ 2008, \pasj, 60, 251 
\bibitem[Jaffe et al.(2004)]{Jaffe2004} Jaffe, W., Meisenheimer, K., R{\"o}ttgering, H.~J.~A., et al.\ 2004, \nat, 429, 47 
\bibitem[Kallman et al.(2004)]{xstar} Kallman, T.~R., Palmeri, P., Bautista, M.~A., Mendoza, C., \& Krolik, J.~H.\ 2004, \apjs, 155, 675 
\bibitem[Kokubun et al. (2007)]{kokubun}Kokubun, M., et al.\ 2007, \pasj, 59, 53 
\bibitem[Koyama et al.(2007)]{Koyama07} Koyama, K., et al.\ 2007, \pasj, 59, 23 
\bibitem[Kriss et al.(1980)]{Kriss} Kriss, G.~A., Canizares, 
C.~R., \& Ricker, G.~R.\ 1980, \apj, 242, 492 
\bibitem[Lutz et al.(2004)]{Lutz2004} Lutz, D., Maiolino, R., Spoon, H.~W.~W., \& Moorwood, A.~F.~M.\ 2004, \aap, 418, 465 
\bibitem[Magdziarz \& Zdziarski(1995)]{pexrav} Magdziarz, P., \& Zdziarski, A.~A.\ 1995, \mnras, 273, 837 
\bibitem[Maiolino et  al.(2010)]{Maiolino2010} Maiolino, R., Risaliti, G., Salvati, M., et al.\ 2010, \aap, 517, A47 
\bibitem[Malizia et al.(2009)]{Malizia2009} Malizia, A., Stephen, 
J.~B., Bassani, L., et al.\ 2009, \mnras, 399, 944 



\bibitem[Marinucci et al.(2012a)]{Marinucci4945} Marinucci, A., Risaliti, G., Wang, J., et al.\ 2012a, \mnras, 423, L6 
\bibitem[Marinucci et al. (2012b)]{Marinucci4507} Marinucci, A. et al. MNRAS submitted
\bibitem[Matt(2002)]{Matt02} Matt, G.\ 2002, \mnras, 337, 147 
\bibitem[Matt et al.(2004)]{Matt04} Matt, G., Bianchi, S., D'Ammando, F., \& Martocchia, A.\ 2004,
\aap, 421, 473 
\bibitem[Mewe et al.(1985)]{Mewe85} Mewe, R., Gronenschild, E.~H.~B.~M., \& van den Oord, G.~H.~J.\ 1985, \aaps, 62, 197 
\bibitem[Miller et al.(2007)]{Miller766} Miller, L., Turner, T.~J., Reeves, J.~N., et al.\ 2007, \aap, 463, 131 
\bibitem[Miller et  al.(2008)]{Miller08} Miller, L., Turner, T.~J., \& Reeves, J.~N.\ 2008, \aap, 483, 437 
\bibitem[Miller et al.(2010)]{Miller2010} Miller, L., Turner, T.~J., Reeves, J.~N., Lobban, A., Kraemer, S.~B., \& Crenshaw, D.~M.\ 2010, \mnras, 403, 196 
\bibitem[Mitsuda et al.(2007)]{Mitsuda07} Mitsuda, K., et al.\ 2007, \pasj, 59, 1 
\bibitem[Moran et al.(2000)]{Moran} Moran, E.~C., Barth, A.~J., Kay, L.~E., \& Filippenko, A.~V.\ 2000, \apjl, 540, L73 
\bibitem[Murphy \& Yaqoob(2009)]{Mytor} Murphy, K.~D., \& Yaqoob, T.\ 2009, \mnras, 397, 1549 
\bibitem[Nenkova et al.(2002)]{Nenkova2002} Nenkova, M., 
Ivezi{\'c}, {\v Z}., \& Elitzur, M.\ 2002, \apjl, 570, L9 
\bibitem[Nenkova et al.(2008)]{Nenkova2008} Nenkova, M., Sirocky, M.~M., Nikutta, R., Ivezi{\'c}, {\v Z}., \& Elitzur, M.\ 2008, \apj, 685, 160 
\bibitem[Palmeri et al.(2003)]{Palmeri} Palmeri, P., Mendoza, C., Kallman, T.~R., Bautista, M.~A., \& Mel{\'e}ndez, M.\ 2003, \aap, 410, 359 
\bibitem[Piconcelli et al.(2007)]{Piconcelli2007} Piconcelli, E., Bianchi, S., Guainazzi, M., Fiore, F., \& Chiaberge, M.\ 2007, \aap, 466, 855 
\bibitem[Poncelet et al.(2006)]{Poncelet2006} Poncelet, A., Perrin, G., \& Sol, H.\ 2006, \aap, 450, 483 
\bibitem[Poutanen \& Svensson(1996)]{Poutanen} Poutanen, J., \& Svensson, R.\ 1996, \apj, 470, 249 
\bibitem[Puccetti et al.(2007)]{Puccetti} Puccetti, S., Fiore,  F., Risaliti, G., et al.\ 2007, \mnras, 377, 607 
 \bibitem[Reeves et al.(2009)]{Reeves456} Reeves, J.~N., O'Brien,  P.~T., Braito, V., et al.\ 2009, \apj, 701, 493 
\bibitem[Risaliti(2002)]{Risa2002} Risaliti, G.\ 2002, \aap, 386, 379 
\bibitem[Risaliti et al.(2002)]{Risaliti2002} Risaliti, G., Elvis, M., \& Nicastro, F.\ 2002, \apj, 571, 234 
\bibitem[Risaliti et al.(2005)]{Risaliti05} Risaliti, G., Bianchi, S., Matt, G., Baldi, A., Elvis, M., Fabbiano,G., \& Zezas, A.\ 2005,  \apjl, 630, L129 
\bibitem[Risaliti et al.(2007)]{Risaliti07} Risaliti, G., Elvis,  M., Fabbiano, G., et al.\ 2007, \apjl, 659, L111 
\bibitem[Risaliti et al.(2009)]{Risa2009a} Risaliti, G., et al.\ 2009, \mnras, 393, L1 
\bibitem[Risaliti et al.(2010)]{Risaliti10} Risaliti, G., Elvis,  M., Bianchi, S., \& Matt, G.\ 2010, \mnras, 406, L20 
\bibitem[Risaliti et al.(2011)]{Risaliti11} Risaliti, G., Nardini,  E., Salvati, M., et al.\ 2011, \mnras, 410, 1027 
\bibitem[Rivers et al.(2011)]{Rivers} Rivers, E., Markowitz, A., \& Rothschild, R.\ 2011, \apjs, 193, 3
 \bibitem[Shu et al.(2011)]{Shu2011} Shu, X.~W., Yaqoob, T., \& Wang, J.~X.\ 2011, \apj, 738, 147 
\bibitem[Spergel et al.(2003)]{Spergel2003} Spergel, D.~N., et al.\ 2003, \apjs, 148, 175 
\bibitem[Takahashi et al. (2007)]{Takahashi}Takahashi, T., et  al.\ 2007, \pasj, 59, 35 

\bibitem[Tatum et al. (2012)]{Tatum}Tatum, M., Turner, T.J., Miller, L., Reeves, J.N., 2012, ApJ, submitted
\bibitem[Tueller et al.(2010)]{Tueller10} Tueller, J., et al.\ 2010, \apjs, 186, 378 
\bibitem[Turner et al.(1997)]{Turner1997} Turner, T.~J., George, I.~M., Nandra, K., \& Mushotzky, R.~F.\ 1997, \apj, 488, 164 
\bibitem[Turner et al.(2000)]{Turner2000} Turner, T.~J., Perola, 
G.~C., Fiore, F., et al.\ 2000, \apj, 531, 245 
  
  \bibitem[Turner et  al.(2007)]{Turner766} Turner, T.~J., Miller, L., Reeves, J.~N., \& Kraemer, S.~B.\ 2007, \aap, 475, 121 
\bibitem[Turner et al.(2008)]{Turner2008} Turner, T.~J., Reeves, J.~N., Kraemer, S.~B., \& Miller, L.\ 2008, \aap, 483, 161 
\bibitem[Turner \& Miller(2009)]{Turner2009} Turner, T.~J., \& Miller, L.\ 2009, \aapr, 17, 47 
\bibitem[Turner et al.(2009)]{Turner0419} Turner, T.~J., Miller, L., Kraemer, S.~B., Reeves, J.~N., \& Pounds, K.~A.\ 2009, \apj, 698, 99 
\bibitem[Turner et al.(2012)]{Turner2012} Turner, T.~J., Miller,  L.,  \& Tatum, M.\ 2012, American Institute of Physics Conference Series, 1427, 165 
\bibitem[Urry \& Padovani(1995)]{Urry1995} Urry, C.~M., \& Padovani, P.\ 1995, \pasp, 107, 803 
\bibitem[Winter et al.(2009)]{Winter09} Winter, L.~M., Mushotzky, R.~F., Reynolds, C.~S., \& Tueller, J.\ 2009, \apj, 690, 1322 
\bibitem[Yaqoob \& Murphy(2011)]{CSYaqoob} Yaqoob, T., \& Murphy, K.~D.\ 2011, \mnras, 412, 277 
\bibitem[Yaqoob(2012)]{Yaqoob4945} Yaqoob, T.\ 2012, \mnras, 423, 
3360 




\end{thebibliography}
